\documentclass[graybox]{svmult}

\usepackage{mathptmx}       
\usepackage{helvet}         
\usepackage{courier}        
\usepackage{type1cm}        
\usepackage{makeidx}         
\usepackage{graphicx}        
\usepackage{multicol}        
\usepackage[bottom]{footmisc}
\usepackage{bm}

\makeindex             

\begin{document}

\title*{Agent-based modeling of zapping behavior of viewers, 
television commercial allocation, 
and advertisement markets}
\titlerunning{Agent-based modeling of TV commercial markets}
\author{Hiroyuki Kyan and Jun-ichi Inoue}
\institute{Hiroyuki Kyan \at Graduate School of Information Science and Technology, 
Hokkaido University, N14-W-9, Kita-ku, Sapporo 060-0814, Japan \email{kyan@complex.ist.hokudai.ac.jp}
\and Jun-ichi Inoue \at Graduate School of Information Science and Technology,
Hokkaido University, N14-W-9, Kita-ku, Sapporo 060-0814, Japan \email{jinoue@cb4.so-net.ne.jp}}
%
%
\maketitle

\abstract{
We propose a simple probabilistic model of zapping behavior of television viewers.  
Our model might be regarded as a `theoretical platform' to investigate the human collective behavior 
in the macroscopic scale through the zapping action of each viewer at the microscopic level. 
The stochastic process of audience measurements as macroscopic quantities such as television 
program rating point or the so-called gross rating point (GRP for short) are reconstructed using the microscopic modeling 
of each viewer's decision making. 
Assuming that each viewer decides the television station to watch by means of three factors, namely, physical constraints 
on television controllers, exogenous information such as advertisement of program by television station, 
and endogenous information given by `word-of-mouth communication' through 
the past market history, we shall construct an aggregation probability of Gibbs-Boltzmann-type with the energy function. 
We discuss the possibility for the ingredients of the model system to exhibit the collective behavior due to not exogenous but endogenous information.  
}

\section{Introduction}
Individual human behaviour is actually an attractive topic 
for both scientists and engineers, and in particular for psychologists,  
however, it is still extremely difficult for us to deal with  
the problem by making use of scientifically reliable investigation. 
In fact, it seems to be somehow an `extraordinary material' for 
exact scientists such as physicists to tackle as their own major. 
This is because there exists quite large sample-to-sample fluctuation 
in the observation of individual behaviour.  
Namely, one cannot overcome the difficulties caused by individual variation to 
find the universal fact in the behaviour. 

On the other hand, in our human `collective' behaviour instead of individual, 
we sometimes observe several universal phenomena which seem to be suitable materials (`many-body systems') for computer scientists  
to figure out the phenomena through agent-based simulations. 
In fact,  collective behaviour of interacting agents such as flying birds, moving insects or 
swimming fishes shows highly non-trivial properties. 
The so-called {\it BOIDS} (an algorithm for artificial life by simulated flocks) realizes the collective behaviour of animal flocks 
 by taking into account only a few simple rules for each 
 interacting `intelligent' agent in computer \cite{Reynolds, Makiguchi}. 

Human collective behavior in macroscopic scale 
is induced both exogenously and endogenously by the result of decision 
making of each human being at the microscopic level. 
To make out the essential mechanism of emergence phenomena, 
we should describe the system by 
mathematically tractable models which should be constructed as simple as possible. 

It is now manifest that there exist quite a lot of suitable examples around us  for such collective behavior 
emerged by our individual decision making. 
Among them, 
the relationship between the zapping actions of viewers and 
the arrangement of programs or commercials is a remarkably reasonable example. 
\begin{figure}[ht]
\begin{center}
\includegraphics[width=10cm]{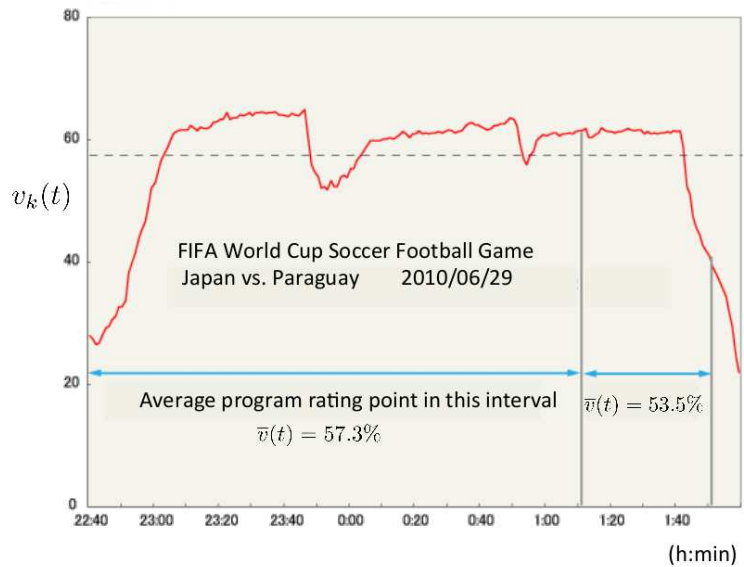}
\end{center}
\caption{\footnotesize 
A typical behavior of program rating point. 
The data set is 
provided by Video Research Ltd. 
(http://www.videor.co.jp/eng/index.html)
}
\label{fig:fg_video}
\end{figure}
\mbox{}

As an example, in Fig. \ref{fig:fg_video}, 
we plot the empirical data for the rating point 
for the TV program of 
FIFA World Cup Soccer Football Game, 
Japan vs. Paraguay which was broadcasted in a Japanese 
television 
station on 29th June 2010. 
From this plot, we clearly observe two large valleys at 
23:50 and 0:50. 
The first valley corresponds to the 
interval between the 1st half and the 2nd half of the game, 
whereas the second valley corresponds to 
the interval between 2nd half of the game and 
penalty shoot-out. 
In these intervals, a huge amount of viewers 
changed the channel to the other stations, and 
one can naturally assume that the program rating point remarkably 
dropped in these intervals. 
Hence, it might be possible to estimate the microscopic viewers' decision 
makings from the macroscopic behavior of the time series such as 
the above program rating point.

Commercials usually  
being broadcasted on 
the television are now well-established as powerful and effective 
tools for sponsors 
to make viewers  
recognize 
their commodities or 
leading brand of the product or service. 
From the viewpoint of television 
stations, 
the commercial 
is quite important to make
a profit as advertising revenue. 
However, at the same time, 
each television station has 
their own wishes to gather viewers of 
their program without any interruption due to 
the commercial because 
the commercial time is also a good chance for the viewers 
to change the channel to check the other programs which 
have been broadcasted from the other rival stations. 
On the other hand, 
the sponsors seek to maximize the so-called {\it contact time} with the viewers 
which has a meaning of duration of their watching the commercials for sponsors' products or survives. 
To satisfy these two somehow distinct demands for 
the television station and sponsors, 
the best possible strategy is to lead the viewers not to 
zap to the other channels during their program. 
However, it is very hard requirement  because we usually 
desire to check the other channels in the hope that 
we might encounter much more attractive programs in that time interval. 

As the zapping action of viewers 
is strongly dependent on the 
preference of the viewers themselves in the first place, 
it seems to be very difficult problem for us 
to understand the phenomena 
by using exact scientific manner. 
However, if we consider the `ensemble' of viewers to 
figure out the statistical properties of 
their collective behavior, 
the agent-based simulation might be an effective tool. 
Moreover, from the viewpoint of human engineering, 
there might exist some 
suitable channel locations  
for a specific television station 
in the sense that 
it is much easier for viewers to zap the channel to arrive 
as a man-machine interface. 

With these mathematical and engineering motivations in mind, here 
we shall propose a simple mathematical model 
for zapping process of viewers. 
Our model system is numerically investigated 
by means of agent-based simulations. 
We evaluate several useful quantities 
such as {\it television program rating point} or 
{\it gross rating point (GRP for short)} from the microscopic 
description of the decision making by each viewer. 
Our approach enables us to 
investigate the television commercial market 
extensively like financial markets \cite{Ibuki}. 

This paper is organized as follows. 
In the next section \ref{sec:sec2}, 
we introduce our mathematical model system and 
several relevant quantities 
such as the program rating point or 
the GRP. 
In section \ref{sec:sec3}, 
we 
clearly introduce Ising spin-like variable 
which 
denotes the time-dependent microscopic state of 
a single viewer, a television station for a 
given arrangement of programs and commercials. 
In the next section \ref{sec:sec4}, 
we show that the macroscopic quantities 
such as program rating point or the GRP 
are calculated in terms of the microscopic variables 
which is introduced in the previous section \ref{sec:sec3}. 
In section \ref{sec:sec5}, 
the energy function 
which specifies the decision making of each viewer 
is introduced explicitly. 
The energy function 
consists of three distinct parts, 
namely, 
a physical constraint on the controller, partial energies by 
exogenous and endogenous information. 
The exogenous part comes from 
advertisement of the program by the television station, 
whereas the endogenous part is regarded as 
the influence by the average program rating point
on the past history of the market. 
By using the maximum entropy principle under several constraints, 
we derive the aggregation probability of viewers 
as a Gibbs-Boltzmann form. 
In section \ref{sec:sec6}, 
we show our preliminary results obtained by computer simulations. 
We also consider the `adaptive location' of commercial advertisements  
in section \ref{sec:adapt}. 
In this section, 
we also consider the effects of 
the so-called {\it Yamaba CMs}, 
which are the successive CMs broadcasted intensively at the climax 
of the program, on 
the program rating points to the advertisement measurements. 
The last section is devoted to the concluding remarks. 
\section{The model system}
\label{sec:sec2}
We first introduce our model system of 
zapping process and submission procedure of 
each commercial into the public through the television programs. 
We will eventually find that these two probabilistic processes turn out to be our 
effective television 
commercial markets.  

As long as we surveyed carefully, 
quite a lot of empirical studies on the effect of commercials on consumers' 
interests have been done, however, 
up to now there are only a few theoretical studies concerning the present research topic 
to be addressed. 
For instance, Siddarth and Chattopadhyay \cite{Siddarth} (see also the references therein)
introduced a probabilistic model of 
zapping process, however, 
they mainly focused on the individual zapping action, 
and our concept of `collective behavior' was 
not taken into account. 
Ohnishi {\it et. al.} \cite{Ikai}  tried to 
solve the optimal arrangement problem of television commercials for a 
given set of constraints in the literature of linear programming. 
Hence, it should be stressed that the goals of their papers are completely different from ours. 
\subsection{Agents and macroscopic quantities}
\begin{figure}[ht]
\begin{center}
\includegraphics[width=10cm]{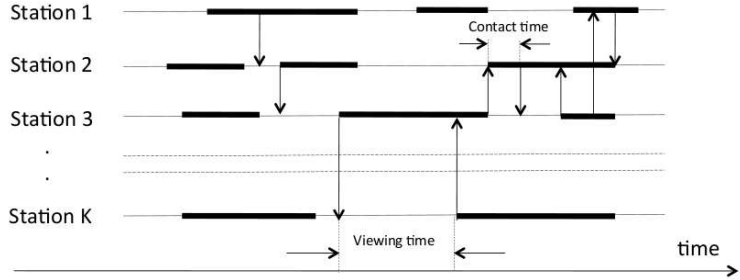}
\end{center}
\caption{\footnotesize 
Two-types of agents in our model systems. 
The thick line segments denote 
the periods of commercial, whereas 
the thin line segments stand for the 
program intervals. 
The set of solid arrows describes a typical 
trajectory of viewer's zapping processes.  
`Contact time' and `viewing time' are clearly defined as 
intervals for which the viewer watches the commercials and the programs, 
respectively. 
The duration and the procedures of casting television programs and CMs  
would be modeled by Poisson arrival processes. The detail accounts for them 
will be given in subsections \ref{subsec:C}, \ref{subsec:D} and 
\ref{subsec:E}. 
}
\label{fig:fg0}
\end{figure}

To investigate the stochastic process of 
zapping process and its influence on the television commercial 
markets, we first introduce two distinct agents, 
namely, television stations, each of which is 
specified by the label 
$k=1,\cdots, K$ 
and viewers specified by the index  
$i=1,\cdots, N$. Here it should be noted that $K\ll N$ should hold. 
The relationship between these two distinct agents is  
described schematically in Fig.\ref{fig:fg0}.  
In this figure, 
the thick line segments denote 
the period of commercial, whereas 
the thin line segments stand for the 
program intervals. 
The set of solid arrows describes a typical 
trajectory of viewer's zapping process. 
Then, the {\it (instant) program rating point} for 
the station $k$ at time $t$ is given by 
\begin{equation}
v_{k}(t) = 
\frac{N_{k}(t)}{N}
\label{eq:vk}
\end{equation}
where 
$N_{k}(t)$ is the number of 
viewers 
who actually watch the television program being 
broadcasted on the channel (the television station) $k$ at time $t$. 

On the other hand, 
the time-slots for commercials are 
traded between 
the station and sponsors 
through the quantity, the so-called  
{\it gross rating point (GRP)} which is 
defined by 
\begin{equation}
GRP^{(n)}_{k} = 
\frac{\theta_{k}^{(n)}}{T}
\sum_{t=1}^{T}
v_{k}(t) 
\label{eq:GRP}
\end{equation}
where 
$T$ denotes total observation 
time, for instance, say $T=600$ minutes,  for evaluating the program rating point.  
$\theta_{k}^{(n)}$ 
stands for the {\it average contact time} for 
viewers  who are watching 
the commercial of the sponsor 
$n$ being 
broadcasted on the station $k$ during 
the interval $T$. 
Namely, the average GRP is defined by the 
product of 
the average program rating point and 
the average contact time. 
It should be noted that 
the equation (\ref{eq:GRP}) is defined 
as the average over the observation time 
$T$. Hence, if one seeks for 
the total GRP of the station $k$ over the observation time 
$T$, 
it should be given as 
$T \times GRP^{(n)}_{k}$. 
Therefore, 
the average GRP 
for the sponsor 
$n$ during 
the observation interval 
$T$ is apparently evaluated by the quantity: 
$GRP^{(n)} = 
(1/K)
\sum_{k=1}^{K}
GRP_{k}^{(n)}$ 
when one assumes that 
the sponsor $n$ 
asked all stations to broadcast their commercials 
through grand waves. 
\subsection{Zapping as a `stochastic process'}
\begin{figure}[ht]
\begin{center}
\includegraphics[width=5cm]{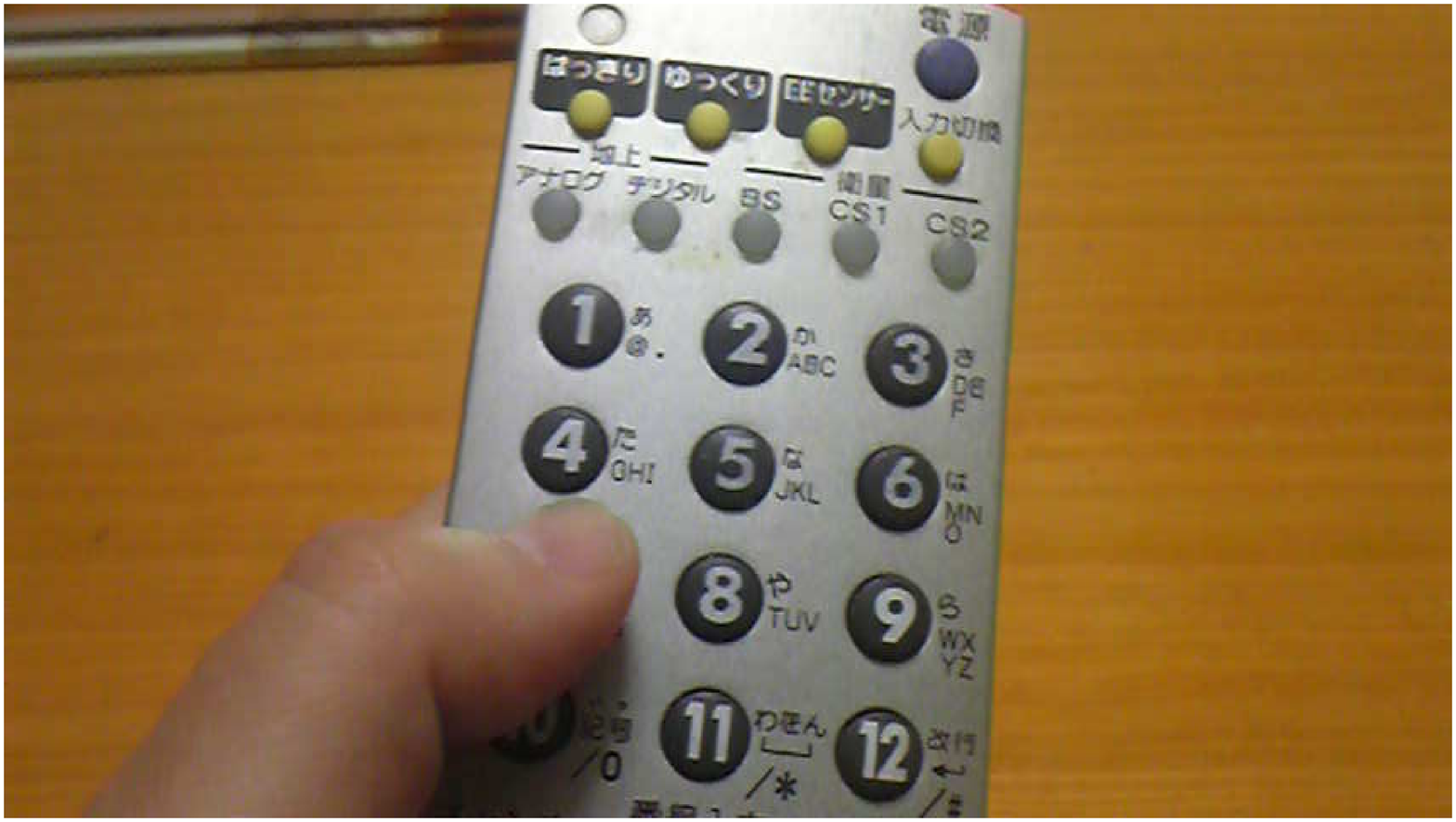} \hspace{0.5cm}
\includegraphics[width=5cm]{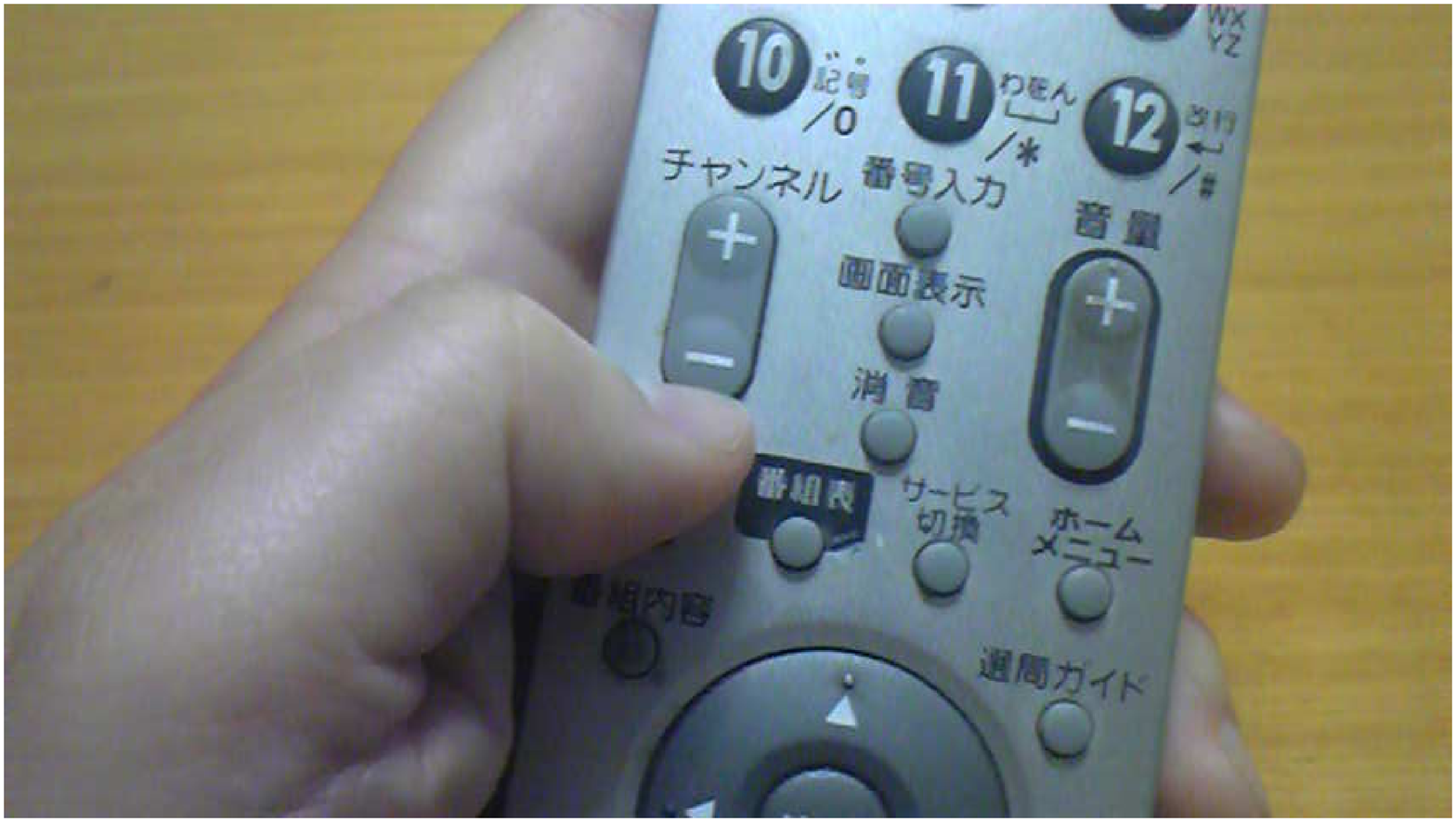}\\
\mbox{}\vspace{0.1cm}\\
\includegraphics[width=5cm]{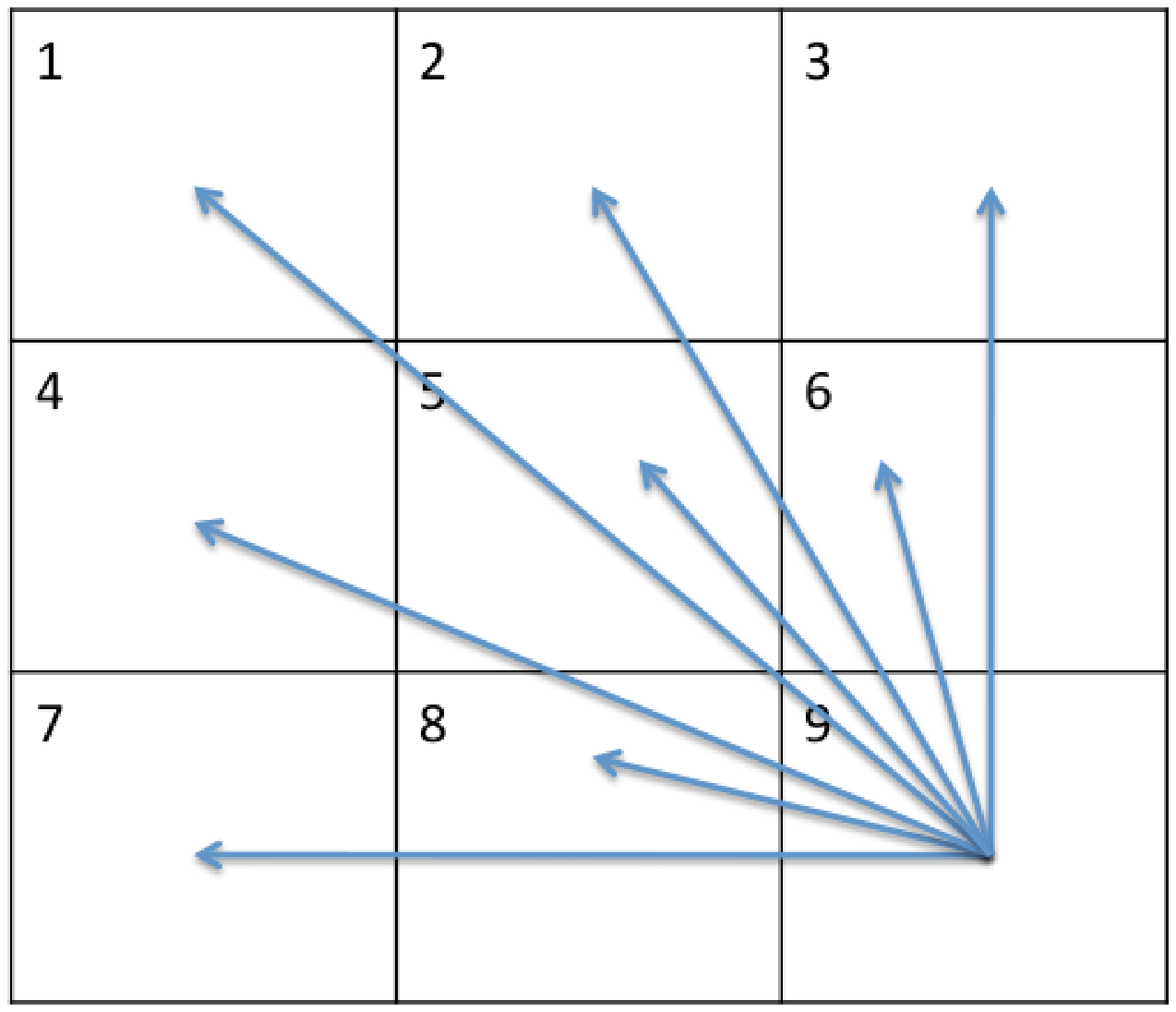} \hspace{0.3cm}
\includegraphics[width=5cm]{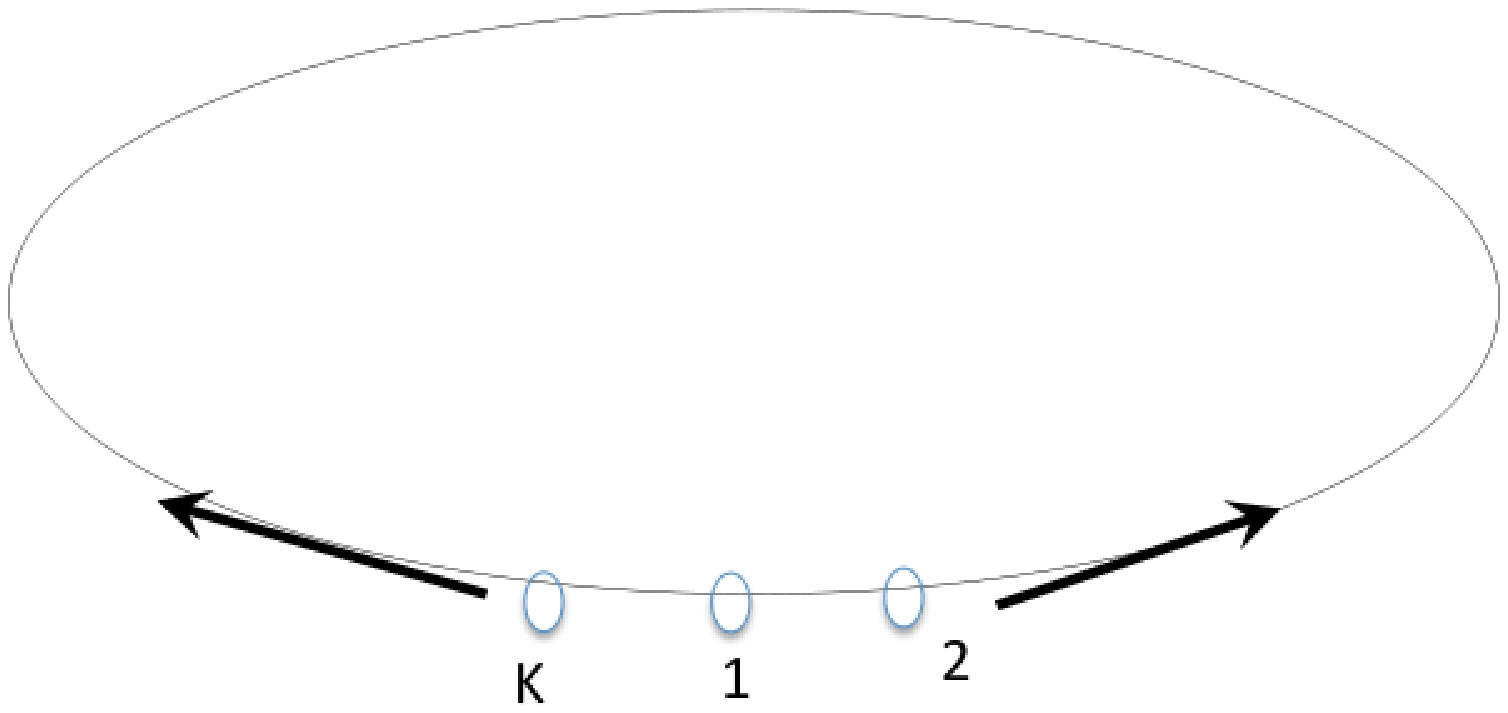}
\end{center}
\caption{\footnotesize
A typical television controller dealt with in this paper. 
The left panel shows 
`lattice-type' channel, whereas the right 
is a typical channel of `ring-type'. 
Each channel location is modeled as shown in 
the corresponding cartoons in the lower two panels.
Namely, the `lattice-type' (left) and 
the `ring-type' (right) for 
the modeling of 
the buttons 
on the television controller.  
Apparently, there exist a geometrical constraint on the latter type. 
 }
\label{fig:fg2}
\end{figure}

We next model the zapping process of viewers. 
One should keep in mind 
that here we consider the controller 
shown in Fig. \ref{fig:fg2}. 
We should notice that 
(at least in Japan) there exist two types of 
channel locations on 
the controller, namely, 
`lattice-type' and 
`ring-type' as shown in Fig. \ref{fig:fg2}. 
For the `lattice-type', 
each button 
corresponding to each station 
is located on the vertex in the two-dimensional square lattice. 
Therefore, for the case of 
$K=9$ stations (channels) for example 
(See the lower left panel in Fig. \ref{fig:fg2}), 
the viewers can change the channel 
from an arbitrary station 
$k$ to 
the other station 
$l (\neq k)$, and there is no geometrical constraint for users (viewers). 
Thus, 
we naturally define 
the transition 
probability $P(l|k)$ which 
is the probability that 
a user change the channel from 
station $k$ to $l(\neq k)$. 
Taking into account the normalization of 
the probabilities, 
we have $
\sum_{l \neq k}^{K}
P(l|k) =1,\,\,
k=1,\cdots,K$. 
The simplest choice of 
the modeling of the transition 
probability satisfying the above constraint is 
apparently 
the uniform one and 
it is written by  
\begin{equation}
P(l|k) = 
\frac{1}{K-1},\,\,\,
k,l (\neq k)=1,\cdots,K.
\label{eq:random_selection}
\end{equation}

On the other hand, 
for the `ring-type' which is shown in 
the lower right panel of Fig. \ref{fig:fg2}, 
we have a geometrical constraint 
$P(l=k \pm 1|k)>0,\,\,P(l \neq k \pm 1|k)=0$. 
From the normalization condition of the probabilities, we also have 
another type of constraint 
$P(l=k-1|k)+P(l=k+1|k)=1$ 
for $k,l(\neq k)=1,\cdots,K$. 
The simplest choice of 
the probability which 
satisfies 
the above constraints is given by 
$P(l=k-1|k)=P(l=k+1|k)=1/2$ for 
$k,l(\neq k)=1,\cdots,K$. 
Namely, the viewers can change from the current channel to the nearest neighboring two stations. 
\subsection{The duration of viewer's stay}
\label{subsec:C}
After changing the channel stochastically to the other rival stations,
the viewer eventually stops to zap and stays the channel to watch 
the program if he or she 
is interested in it. 
Therefore, we should construct 
the probabilistic model of 
the length of viewer's stay in both programs and commercials appropriately. 
In this paper, we assume that 
the lengths of the viewer $i$'s stay in  the programs $\tau_{\rm on}$ 
(`on' is used as an abbreviation for `on air') 
and commercials $\tau_{\rm cm}$ are given 
as `snapshots' from the distributions: 
\begin{equation}
P^{(i)}(\tau_{\rm on,cm}) = 
a_{\tau_{\rm on,cm}}^{(i)}\,{\rm e}^{
-\tau_{\rm on,cm}/a_{\tau_{\rm on,cm}}^{(i)}
},\,\,\,i=1,\cdots, N
\end{equation}
where 
$a_{\tau_{\rm on,cm}}^{(i)}$ 
stands for the `relaxation time' of 
the viewer $i$ for 
the program and the commercial, 
respectively. 
Of course, 
$\tau_{\rm on,cm}$ fluctuates 
from person to person, 
hence, here we assume that 
$a_{\tau_{\rm on,cm}}^{(i)}$ might follow
\begin{equation}
a_{\tau_{\rm on,cm}}^{(i)} = c_{\rm on,cm} + \delta,\,\,
P(\delta) =\mathcal{N}(0,\sigma^{2}).
\end{equation} 
Namely, the relaxation 
times for 
the program and commercial 
fluctuate 
around the typical value $c_{\rm on,cm}$ 
by a white Gaussian noise 
with mean zero and variance 
$\sigma^{2}$. 
Obviously, 
for ordinary viewers, 
$c_{\rm on}>c_{\rm cm}$ should 
be satisfied. 
We should notice that the above choice of 
the length of 
viewer's stay is independent on 
the station, program or sponsor. 
For instance, 
the length of viewer's stay in commercials 
might be changed 
according to the combination of 
commercials of different kinds of sponsors. 
However, if one needs, we can modify 
the model by taking into account the corresponding empirical data. 
\subsection{The process of casting commercials}
\label{subsec:D}
Here we make a model of casting commercials by television stations. 
To make a simple model, 
we 
specify each sponsor by the label $n=1,\cdots,M$ and 
introduce microscopic variables $l_{k}(t)$ 
as follows. 
\begin{equation}
l_{k}(t)=
\left\{
\begin{array}{ll}
n \in \{1,\cdots, M\} & \mbox{(The station $k$ casts a CM of sponsor $n$ at time $t$)} \\
0 & \mbox{(The station $k$ casts a program at $t$)}
\end{array}
\right.
\end{equation}
Hence, 
if one obtains 
$l_{1}(1)=l_{1}(2)=\cdots=l_{1}(10)=3$ and  $l_{1}(11)=0$, 
then we conclude that 
the station $k=1$ casted 
the commercial of the sponsor $n=3$ from $t=1$ to $t=10$, 
and after this commercial period, 
the station $k$ resumed the program at the next step $t=11$. 
On the other hand, 
if we observe 
$l_{1}(1)=l_{1}(2)=\cdots=l_{1}(10)=3$ and $l_{1}(11)=2$, 
we easily recognize 
that 
the station $k=1$ casted 
the commercial of the sponsor $n=3$ from $t=1$ to $t=10$, 
and after this commercial period, 
the same station $k=1$ 
casted the commercial of the sponsor $n=2$ at the next step $t=11$. 
Therefore,  for a 
given sequence of 
variables $l_{k}(t),\,k=1,\cdots,K$ 
for observation period $t=1,\cdots, T$, 
the possible patterns being broadcasted by all television stations 
are completely determined. 
Of course, we 
set $l_{k}(t),\,\,k=1,\cdots,K$ 
to the artificial values in our 
computer simulations, 
however, 
empirical evidence might help us to 
choose them. 
\subsection{Arrangement of programs and CMs}
\label{subsec:E}
In following, 
we shall explain how each television station 
submits the commercials to appropriate time-slots of their broadcasting. 
First of all, 
we set $l_{k}(0)=0$ for all stations 
$k=1,\cdots,K$. Namely, we assume that 
all stations start their broadcasting 
from their own  program instead of any commercials of their sponsors.
Then, for an arbitrary $k$-th station, 
the duration $t_{\rm on}$ between the starting and the ending points of each section in the program 
is generated by the exponential distribution $\sim {\rm e}^{-t_{\rm on}/L_{\rm on}}$. 
Thus, from the definition, we should set $l_{k}(0)=\cdots=l_{k}({t_{\rm on}})=0$ for the resulting 
$t_{\rm on}$. 
We next choose a sponsor among 
the $M$-candidates by sampling from a uniform distribution in $[1,M]$. 
For the selected sponsor, say, $n \in \{1,M\}$, 
the duration of their commercial $t_{\rm cm}$ is 
determined by a snapshot of the exponential 
distribution $\sim {\rm e}^{-t_{\rm cm}/L_{\rm cm}}$. 
Thus, from the definition,  we set 
$l_{k}(t_{\rm on}+1)=\cdots=
l_{k}(t_{\rm on}+1+t_{\rm cm})=n$ for 
the given $t_{\rm cm}$ and  $t_{\rm on}$. 
We repeat the above procedure 
from $t=0$ to $t=T-1$ for 
all stations $k=1,\cdots,K$. 
Apparently, we should choose these 
two relaxation times 
$L_{\rm on},L_{\rm cm}$ so 
as to satisfy $L_{\rm on} \gg L_{\rm cm}$. 
After this procedure, we obtain 
the realization of combinations of `thick' (CMs) and `thin' (television programs) lines 
as shown in Fig. \ref{fig:fg0}. 
\section{Observation procedure}
\label{sec:sec3}
In the above section \ref{sec:sec2}, we introduced the model system. 
To figure out the macroscopic behavior of the system, we should 
define the observation procedure. 
For this purpose, we first introduce 
microscopic binary (Ising spin-like) variables 
$S_{i,k}^{(l_{k}(t))}(t) \in \{0,1\}$ 
which is defined by 
\begin{equation}
S_{i,k}^{(n)}(t) = 
\left\{
\begin{array}{ll}
1 & \mbox{(The viewer $i$ watches the CM of sponsor $n$ on the station $k$ at time $t$)} \\
0 & \mbox{(The viewer $i$ does not watch the CM of $n$ on the station $k$ at time $t$)} 
\end{array}
\right.
\end{equation}
We should notice that for the case of $l_{k}(t)=0$, 
the Ising variable $S_{i,k}^{(l_{k}(t))}(t)$ takes 
\begin{equation}
S_{i,k}^{(0)}(t) = 
\left\{
\begin{array}{ll}
1 & \mbox{(The viewer $i$ watches the program on the station $k$ at time $t$)} \\
0 & \mbox{(The $i$
does not watch the program on the station $k$ at time $t$)} 
\end{array}
\right.
\end{equation}
Thus, the $K\times N$-matrix 
$\bm{S}^{(n)}(t)$ written by 
\begin{equation}
\bm{S}^{(n)}(t) = 
\left(
\begin{array}{ccc}
S_{1,1}^{(n)}(t) & \cdots & 
S_{1,K}^{(n)}(t) \\
\cdots & \cdots & \cdots \\
\cdots & \cdots & \cdots \\
S_{N,1}^{(n)}(t) & \cdots & 
S_{N,K}^{(n)}(t) 
\end{array}
\right)
\end{equation}
becomes a sparsely coded large-size matrix 
which has only a single non-zero entry in each column.   
On the other hand, by summing up all elements 
in each row, the result, say 
$S_{1,k}^{(n)}+ \cdots+S_{N,k}^{(n)}$ 
denotes the number of viewers who watch the commercial 
$n$ on the station $k$ at time $t$. 
Hence, if the number of sorts of commercials (sponsors) $M$ ($n \in \{1,\cdots,M\}$) is 
quite large, the element should satisfy $S_{1,k}^{(n)}+ \cdots+S_{N,k}^{(n)} \ll N$ 
(Actually, it is a rare event that an extensive number of viewers watch the same commercial 
on the station $k$ at time $t$)  
and this means that  the matrix 
$\bm{S}^{(n)}(t)$ is a sparse large-size matrix.  
It should be noted that 
macroscopic quantities such as 
program rating point or the GRP are constructed in terms of the Ising variables 
$S_{i,k}^{(l_{k}(t))}(t) \in \{0,1\}$. 

For the Ising variables 
$S_{i,k}^{(l_{k}(t))}(t) \in \{0,1\}$, 
besides we already mentioned above, 
there might exist several constraints to be satisfied. 
To begin with, as the system has $N$-viewers, 
the condition 
$\sum_{l_{k}(t)=0}^{M}
\sum_{k=1}^{K}
\sum_{i=1}^{N}S_{i,k}^{(l_{k}(t))}(t)=N$
should be naturally satisfied. 
On the other hand, 
assuming that 
each viewer  $i=1,\cdots,N$ 
watches the television without any interruption 
during the observation time $T$, 
we immediately have 
$\sum_{t=1}^{T}
\sum_{l_{k}(t)=0}^{M}\sum_{k=1}^{K}
S_{i,k}^{(l_{k}(t))}(t) = T,\,\, i=1,\cdots,N$. 
It should bear in mind that the viewer $i$ might 
watch the program or commercial brought by 
one of the $M$-sponsors, hence, 
we obtain the condition
$ \sum_{l_{k}(t)=0}^{M}
\sum_{k=1}^{K}S_{i,k}^{(l_{k}(t))}(t)=1,\,\, i=1,\cdots,N,\,\,
t=1,\cdots,T$. 

These conditions might help us to check the validity of programming codes and 
numerical results. 
\section{Micro-descriptions of macro-quantities}
\label{sec:sec4}
In this section, we explain 
how one describes the relevant macroscopic quantities 
such as average program rating point or the GRP 
by means of a set of microscopic Ising variables 
$\{S_{i,k}^{(l_{k}(t))}(t)\}$ which was introduced in the 
previous section. 
\subsection{Instant and average program rating points}
Here we should notice that 
the number of viewers who are watching 
the program or commercials being 
broadcasted on the station  $k \,(= 1,\cdots, K)$, 
namely, $N_{k}(t)$ 
is now easily rewritten in terms of 
the Ising variables at the microscopic level as 
$N_{k}(t) = 
\sum_{i=1}^{N}
\sum_{l_{k}(t)=0}^{M}
S_{i,k}^{(l_{k}(t))}(t)$. 
Hence, from equation (\ref{eq:vk}), 
the instant program rating point of 
the station $k$ at time $t$, 
that is $v_{k}(t)$, 
is given explicitly as 
\begin{equation}
v_{k}(t) = 
\frac{N_{k}(t)}{N}= 
\frac{1}{N}
\sum_{i=1}^{N}
\sum_{l_{k}(t)=0}^{M}
S_{i,k}^{(l_{k}(t))}(t). 
\label{eq:rate1}
\end{equation}
On the other hand, 
the average program rating point 
of the station $k$, 
namely, 
$\overline{v}_{k}$ is 
evaluated as 
$\overline{v}_{k} = 
(1/T)\sum_{t=1}^{T}v_{k}(t)$. 
\subsection{Contact time and cumulative GRP}
We are confirmed that the contact time which was already introduced in 
the previous section \ref{sec:sec2} 
is now calculated in terms of 
$\{S_{i,k}^{(l_{k}(t))}(t)\}$ as follows. 
The contact time of the viewer $i$ with 
the commercial of the sponsor $n$ is written as 
\begin{equation}
\theta_{i,k}^{(n)}=
(1/T)
\sum_{t=1}^{T}
\sum_{l_{k}(t)=0}^{M}
\delta_{l_{k}(t),n}
S_{i,k}^{(l_{k}(t))}(t)
\end{equation}
where 
$\delta_{a,b}$  denotes the Kronecker's delta. 
It should be noted that 
we scaled the contact time 
over the observation time $T$ by 
$1/T$ so as to 
make the quantity the $T$-independent value. 
Hence, 
the average contact time of all 
viewers who watch the commercial of sponsor 
$n$ being broadcasted on the station $k$ 
is determined by 
$\theta_{k}^{(n)} \equiv  
(1/N)
\sum_{i=1}^{N}
\theta_{i,k}^{(n)}$. 
Thus, the cumulative GRP is 
obtained from the definition 
(\ref{eq:GRP}) as 
\begin{eqnarray}
GRP_{k}^{(n)} & = & 
\frac{1}{T}\sum_{t=1}^{T}v_{k}(t) \times \theta_{k}^{(n)} \nonumber \\
\mbox{} & = & 
\frac{1}{N^{2}T^{2}}
\left(
\sum_{t=1}^{T}
\sum_{i=1}^{N}
\sum_{l_{k}(t)=0}^{M}
S_{i,k}^{(l_{k}(t))}(t)
\right) 
\left(
\sum_{i=1}^{N}
\sum_{t=1}^{T}
\sum_{l_{k}(t)=0}^{M}
\delta_{l_{k}(t),n}
S_{i,k}^{(l_{k}(t))}(t) 
\right). 
\label{eq:rate3}
\end{eqnarray}
From the above our argument, 
we are now confirmed that 
all relevant quantities in 
our model system could be 
calculated in terms of 
the Ising variables $\{S_{i,k}^{(l_{k}(t))}(t)\}$ which describe 
the microscopic state of ingredients in the commercial market. 

However, the matrix 
$\bm{S}^{(n)}(t)$ itself is 
determined by the actual stochastic processes of 
viewer's zapping with arranging the programs and television commercials. 
Therefore, in the next section, we
introduce the energy(cost)-based 
zapping probability  
which contains the random selection (\ref{eq:random_selection}) 
as a special case for the `lattice-type' controller. 
\section{Energy function of zapping process}
\label{sec:sec5}
As we showed in the literature of 
our probabilistic labor market \cite{Chen}, 
it is convenient for us to 
construct the energy function 
to quantify the action of 
each viewer. 
The main issue to be clarify in this study 
is the condition on which concentration (`condensation') of viewers to 
a single television station is occurred due to the endogenous 
information. The same phenomena 
refereed to as informational cascade 
in the financial market is observed by modeling 
of the price return by means of 
magnetization in the Ising model \cite{Ibuki2}. In the financial 
problem, 
the interaction $J_{ij}$ between Ising spins $S_{i}$ and $S_{j}$  corresponds to 
endogenous information, whereas the external magnetic field $h_{i}$ affected on 
the spin $S_{i}$ 
stands for the exogenous information. 
However, in our commercial market, 
these two kinds of information would be described by means of 
a bit different manner. It would be given below. 
\subsection{Physical constraints on television controllers}
For this end, 
let us describe here 
the location of channel for the station 
$k$ on the controller as 
a vertex on the 
two-dimensional square lattice (grid) as $\bm{Z}_{k} \equiv (x_{k},y_{k})$. 
Then, we assume that the channel located on 
the vertex at which the distance from the 
channel $k$ is minimized might be more likely to be selected 
by the viewer who watches the program (or commercial) on the station 
$k$ at the instance $t$. 
In other words, 
the viewer minimizes the energy function 
given by   
$\gamma (\bm{Z}_{k}-\bm{Z}_{l})^{2}\,\,\,(l \neq k)$, 
where we defined the $L_{2}$-norm as the distance 
$(\bm{Z}_{k}-\bm{Z}_{l})^{2} \equiv 
(x_{k}-x_{l})^{2}+
(y_{k}-y_{l})^{2}$. The justification of the above assumption should 
be examined from the viewpoint of human-interface engineering. 
\subsection{Exogenous information}
Making the decision of viewers is affected by the exogenous 
information. For instance, 
several weeks before World Cup qualifying game, 
a specific station $\overline{k}$, which will be permitted to broadcast the game,  
might start to advertise the program of the match. 
Then, a large fraction of 
viewers including a soccer football fan might decide to watch the program at the time. 
Hence, the effect might be taken into account 
by introducing the energy $-\zeta \prod_{\xi=s}^{e}\delta_{t_{\xi},t} \delta_{l, \overline{k}}$ 
where $t_{s}$ denotes the time at which the program 
starts and $t_{e}$ stands for the time of the end. 
Therefore, 
the energy decreases when the viewer watches the 
match of World Cup qualifying during the time for the program, 
namely from $t=t_{s}$ to $t_{e}$ ($\Delta t \equiv t_{e}-t_{s}$: broadcasting hours 
of the program). 
\subsection{Endogenous information}
The collective behavior might be caused by 
exogenous information 
which is corresponding to 
`external field' in the literature of statistical physics. 
However, collective behavior of viewers also could be `self-organized'  
by means of endogenous information. 
To realize the self-organization, 
we might use the moving average of 
the instant program rating 
over the past $L$-steps ($L\ll T$), 
namely, 
\begin{equation}
\langle v_{k}(t) \rangle \equiv 
\frac{1}{L}
\sum_{\rho=t-L}^{t-1}
v_{k}(\rho)
\end{equation}
as the endogenous information. 
Then, we define the `winner channel' which is 
more likely to be selected at time $t$ as 
$\hat{k} = 
\arg\max_{m} \langle v_{m}(t) \rangle$ 
(One might extend it to a much more general form:  
\begin{equation}
\hat{k} = 
\arg\max_{m} \langle v_{m}(r) \rangle
\label{eq:lag}
\end{equation}
 for a given `time lag' $r \,(<t)$). 
Henceforth, we assume that if 
the winner channel $\hat{k}$ 
is selected, a part of total energy 
$-\beta \delta_{l,\hat{k}}$ decreases. 
This factor might cause 
the collective behavior of 
$N$-individual viewers. 
Of course, if one needs, it might be possible for us 
to recast the representation of the winner channel 
$\hat{k}$ by means of 
microscopic Ising variables $\{S_{i,k}^{(l_{k}(t)}(t)\}$. 

Usually, the collective behavior is caused 
by direct interactions (connections) between agents. 
However, nowadays, 
watching television is completely a `personal action' which is dependent on the personal 
preference 
because every person can possess their own television 
due to the wide-spread drop in the price of the television set. 
This means that there is no direct interaction between 
viewers, 
and the collective behavior we expect here 
might be caused 
by some sorts of public information such as 
program rating point in the previous weeks. 
In this sense, 
we are confirmed that the above choice of energy should be naturally accepted. 

Therefore, 
the total energy function 
at time $t$ is defined by 
\begin{equation}
E_{k}(l) \equiv  \gamma (\bm{Z}_{k}-\bm{Z}_{l})^{2}- \zeta \prod_{\xi=s}^{e}\delta_{t_{\xi},t} \delta_{l, \overline{k}}- \beta \delta_{l,\hat{k}} 
\label{eq:energy_lattice}
\end{equation}
with $\hat{k} = 
\arg\max_{m} \langle v_{m}(t) \rangle$,  
where $\beta,\zeta,\gamma \geq 0$ are model parameters to be estimated from the empirical data 
in order to calibrate our model system. 
According to the probabilistic 
labor market which was introduced by one of the present authors, 
we construct the transition 
probability $P(l|k)$ 
as the Gibbs-Boltzmann form by 
solving the optimization problem 
of the functional: 
\begin{equation}
f[P(l|k)]  \equiv 
-\sum_{l \neq k}
P(l|k)\log P(l|k) - \lambda \left\{
\sum_{l \neq k}P(l|k)-1\right\} 
-  \lambda^{'} \left\{
\sum_{l \neq k}E_{k}(l)P(l|k)-E\right\}
\end{equation}
with respect to $P(l|k)$. 
Then, we immediately obtain the solution of the 
optimization problem (variational problem) as  
\begin{eqnarray*}
P(l|k) & = &    
\frac{{\exp}[-E_{k}(l)]}
{\sum_{l \neq k}
{\exp}[-E_{k}(l)]} \nonumber \\
\mbox{} & = & 
\frac{{\exp}[-\gamma (\bm{Z}_{k}-\bm{Z}_{l})^{2}+
\zeta \prod_{\xi=s}^{e}\delta_{t_{\xi},t} \delta_{l, \overline{k}} + 
\beta \delta_{l,\hat{k}}]}
{\sum_{l \neq k}
{\exp}[-\gamma (\bm{Z}_{k}-\bm{Z}_{l})^{2}+
\zeta \prod_{\xi=s}^{e}\delta_{t_{\xi},t} \delta_{l, \overline{k}} + 
\beta \delta_{l,\hat{k}} ]} 
\end{eqnarray*}
where we chose one of the Lagrange multipliers in $f[P(l|k)]$ as $\lambda = \log \{\sum_{l \neq k}P(l|k)\}-1$, and 
another one $-\lambda^{'}$ is set to $1$ for simplicity,   
which has a physical meaning of 
`unit inverse-temperature'. 
We should notice that 
in the `high-temperature limit' $\beta, \zeta, \gamma \to 0$, 
the above probability 
becomes identical to that of 
the random selection (\ref{eq:random_selection}). 
These system parameters should be calibrated by 
the empirical evidence. 

In the above argument, we focused on 
the `lattice-type' controller, 
however, it is easy for us to modify the energy function 
to realize the `ring-type' by replacing 
$\gamma (\bm{Z}_{k}-\bm{Z}_{l})^{2}$ in (\ref{eq:energy_lattice}) by 
$\varepsilon_{l} \equiv -\gamma (\delta_{l,k+1}+\delta_{l,k-1})$, 
namely, 
the energy 
decreases if and only if 
the viewer who is watching 
the channel $k$ moves to the television 
station 
$k-1$ or $k+1$. This modification immediately leads to 
\begin{equation}
P(l|k) =    
\frac{{\exp}[-E_{k}(l)]}
{\sum_{l \neq k}
{\exp}[-E_{k}(l)]} =  
\frac{{\exp}[
-\varepsilon_{l}
+
\zeta \prod_{\xi=s}^{e}\delta_{t_{\xi},t} \delta_{l, \overline{k}} + 
\beta \delta_{l,\hat{k}}]}
{\sum_{l \neq k}
{\exp}[
-\varepsilon_{l}
+
\zeta \prod_{\xi=s}^{e}\delta_{t_{\xi},t} \delta_{l, \overline{k}} + 
\beta \delta_{l,\hat{k}} 
]}.
\end{equation}
We are easily confirmed that 
the transition probability for 
random selection in the `ring-type' 
controller is recovered 
by setting $\zeta=\beta=0$ as 
\[
P(k-1|k) = 
\frac{{\rm e}^{\gamma}}{{\rm e}^{\gamma}+{\rm e}^{\gamma}}=\frac{1}{2}=P(k+1|k) 
\]
and $P(l|k) =0$ for $l \neq k \pm 1$. 

In the next section, we show the results from our limited contributions by computer simulations. 
\section{A preliminary: computer simulations}
\label{sec:sec6}
\begin{figure}[ht]
\begin{center}
\includegraphics[width=10cm]{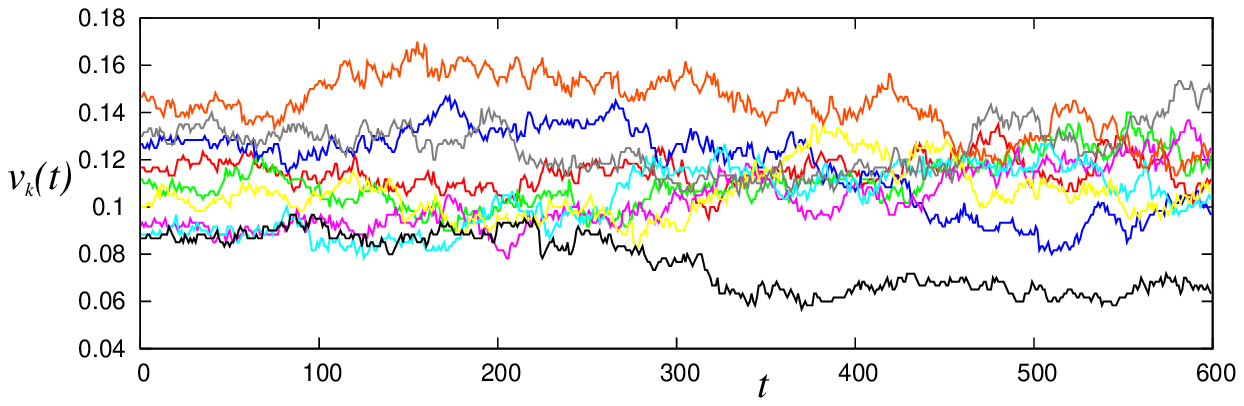} 
\vspace{-0.7cm}
\includegraphics[width=10cm]{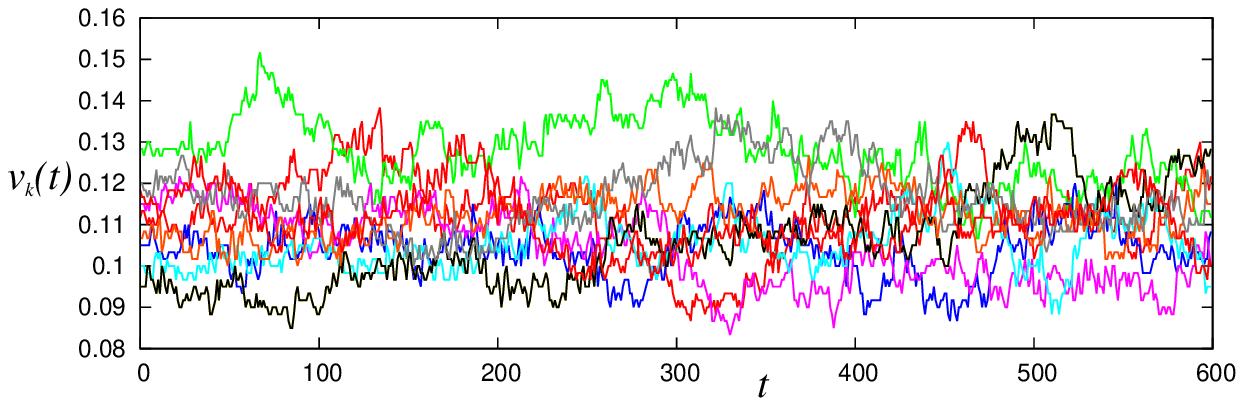}
\end{center}
\caption{\footnotesize 
Typical behavior of 
instant program rating point $v_{k}(t)$ for $k=1,\cdots,K$. 
Here we set $K=9, M=1, N=600$ and $T=600$. 
The parameters appearing in the system are chosen as 
$L_{\rm on}=c_{\rm on}=12$ and $L_{\rm cm}=c_{\rm cm}=4$.  
The upper panel is the result of `lattice-type' with 
$\beta=\zeta=\gamma=0$, whereas 
the lower panel shows the case of `ring-type'. 
We clearly find that the result of ring-type is less volatile 
than that of the lattice-type. }
\label{fig:fg61}
\end{figure}

In this section, we show our preliminary results. 
In Fig. \ref{fig:fg61}, 
we plot the typical behavior of 
instant program rating point $v_{k}(t)$ for $k=1,\cdots,K$. We set $K=9, N=600,M=1$ and $T=600$ 
for the case of the simplest choice $\beta=\zeta=\gamma=0$ leading up to 
(\ref{eq:random_selection}) (`high-temperature limit'). 
This case might correspond to 
the `unconscious zapping' by viewers. 
The parameters appearing in the system are chosen as 
$L_{\rm on}=c_{\rm on}=12$ and $L_{\rm cm}=c_{\rm cm}=4$.  
The upper panel shows the result of `lattice-type' channel location 
on the controller, whereas the lower panel is the result of 
`ring-type'. 
\begin{figure}[ht]
\begin{center}
\includegraphics[width=5.5cm]{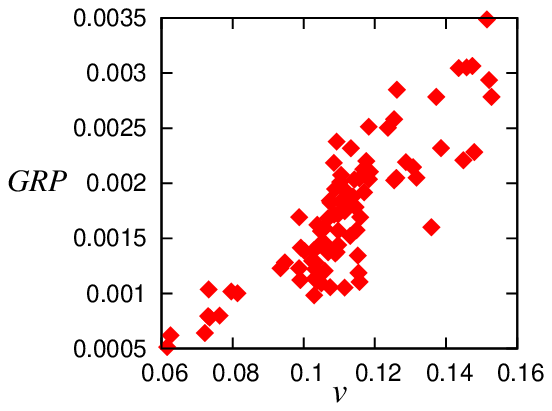} 
\includegraphics[width=5.5cm]{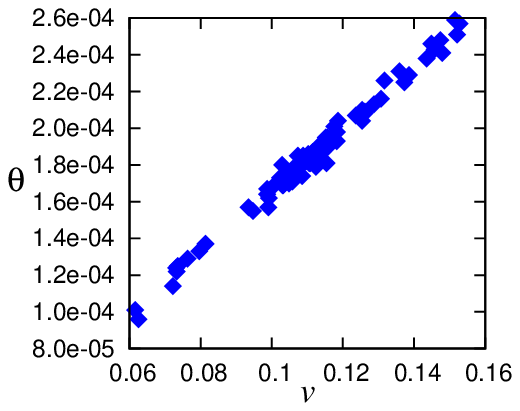} 
\end{center}
\caption{\footnotesize 
The two-dimensional scattered plot for 
the average program rating point:  
$v \equiv (1/K)\sum_{k=1}^{K}\overline{v}_{k}$ and the average 
cumulative GRP: $GRP \equiv (1/K)\sum_{k=1}^{K} GRP_{k}^{(1)}$ for 
the case of `lattice-type' channel location (left). 
The right panel shows  
the scatter plot with respect to the $v$ and 
the effective contact time which is defined 
by $\theta \equiv (1/KT)\sum_{k=1}^{K}\theta_{k}^{(1)}$ (We have only a single sponsor). 
The parameters are set to the same values as in Fig. \ref{fig:fg61}.}
\label{fig:fg62}
\end{figure}

In Fig. \ref{fig:fg62} (left), 
we display the  scattered plot with respect to 
the GRP and the average program rating point for 
the case of `lattice-type' channel location. 
From this figure, we find that 
there exists a remarkable positive correlation between 
these two quantities (the Pearson coefficient is $0.85$). 
This fact is a justification for us to choose  
the GRP as a `market price' for transactions. 
In the right panel of this figure, 
the scattered plot with respect to the GRP and 
the effective contact time defined 
by $(1/KT)\sum_{k=1}^{K}\theta_{k}^{(1)}$ is shown. 
It is clearly found that 
there also exists a positive correlation 
with the Pearson coefficient $0.9914$. 
\begin{figure}[ht]
\begin{center}
\includegraphics[width=9cm]{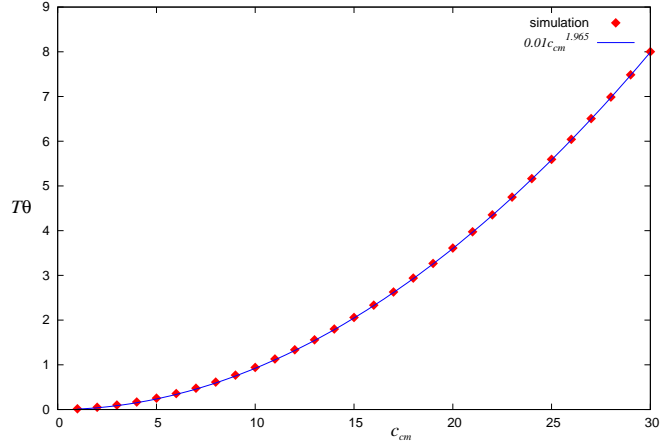}
\end{center}
\caption{\footnotesize 
The $c_{\rm cm}$-dependence of the $T$-scaled effective contact time $T\theta$. 
We set the observation time $T=60$ and 
the other parameters are set to the same values 
as in Fig. \ref{fig:fg61}. 
We also plot the well-fitting curve 
$0.01 c_{\rm cm}^{1.965}$ for eyes' guide. 
}
\label{fig:fg63}
\end{figure}

We also plot the $c_{\rm cm}$-dependence of 
the $T$-scaled effective contact time $T\theta$ in Fig. \ref{fig:fg63}. 
This figure tells us that the frequent zapping actions reduce  
the contact time considerably and it becomes really painful for the sponsors. 
\subsection{Symmetry breaking due to endogenous information}
\begin{figure}[ht]
\begin{center}
\includegraphics[width=10cm]{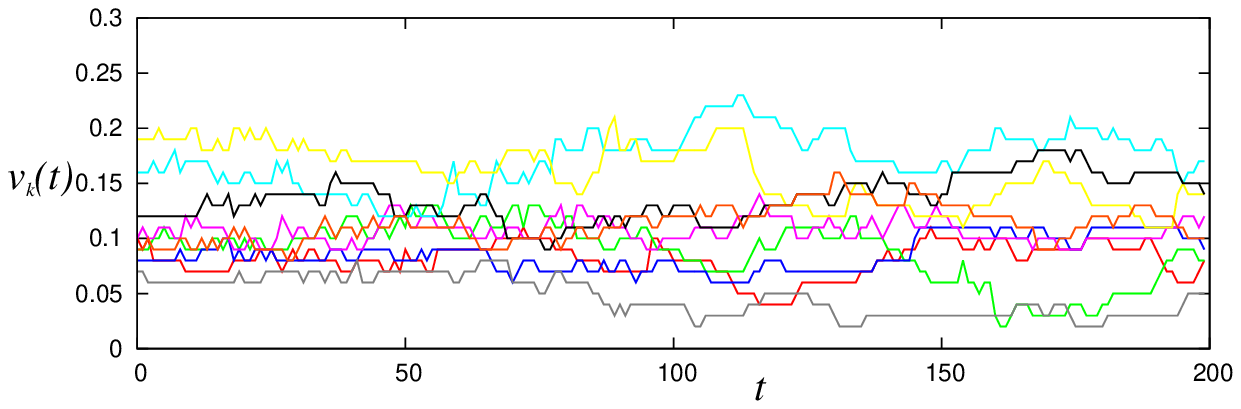} 
\includegraphics[width=10cm]{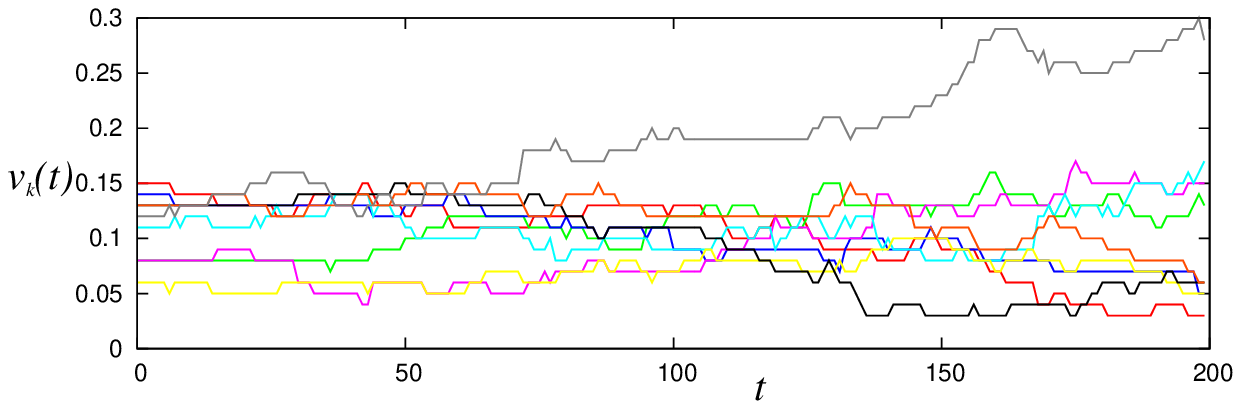}
\end{center}
\caption{\footnotesize 
Typical behavior of 
instant program rating point $v_{k}(t)$ for $k=1,\cdots,K$. 
Here we set $K=9, M=1, N=600$ and $T=200$. 
The parameters which specify the 
energy function of 
`lattice-type' controller are chosen as 
$(\gamma, \zeta,\beta)=(1,0,0)$ (the upper panel) and 
$(\gamma,\zeta,\beta)=(1,0,1.8)$ (the lower panel) with the 
history length $L=5$. 
The parameters appearing in the system are chosen as 
the same values as in Fig. \ref{fig:fg61}.}
\label{fig:fg64}
\end{figure}

We next consider the case 
in which each viewer makes his/her decision 
according to 
the market history, 
namely, 
we choose 
$\gamma=1,\zeta=0$ and set the value of 
$\beta$ to $\beta=0$ and $\beta=1.8$. 
We show the numerical results 
in Fig. \ref{fig:fg64}.  
From this panels, 
we find that the instant program rating point $v_{k}(t)$ for 
a specific television station 
increases so as to become a `monopolistic station' when 
each viewer starts to 
select the station 
according to the market history, 
namely, $\beta  >0$. 
In other words, 
the symmetry of the system 
with respect to 
the program rating point is broken 
as the parameter $\beta$ increases.

\begin{figure}[ht]
\begin{center}
\includegraphics[width=10cm]{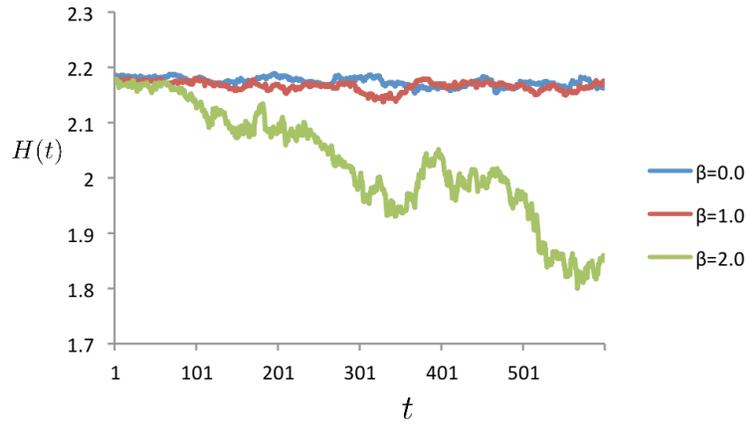} 
\end{center}
\caption{\footnotesize 
A typical behavior of the Shannon's entropy $H(t)$.}
\label{fig:entropy}
\end{figure}

To measure the degree of the `symmetry breaking' in 
the behavior of the instant program rating points more explicitly, 
we introduce the following order parameter: 
\begin{equation}
B(t) \equiv 
\frac{1}{K}
\sum_{k=1}^{K}
\left|
v_{k}(t)-\frac{1}{K}
\right|
\end{equation}
which is defined as 
the cumulative difference 
between $v_{k}$ and 
the value for the `perfect 
equality' $1/K$. 
We plot the $B(t)$ for 
the case of $\beta=0,1$ and $2$. 
For finite $\beta$, the symmetry is apparently broken 
around $t=100$ and 
the system changes from 
{\it symmetric phase} (small $B(t)$) to 
the {\it symmetry breaking phase} (large $B(t)$). 
\begin{figure}[ht]
\begin{center}
\includegraphics[width=9cm]{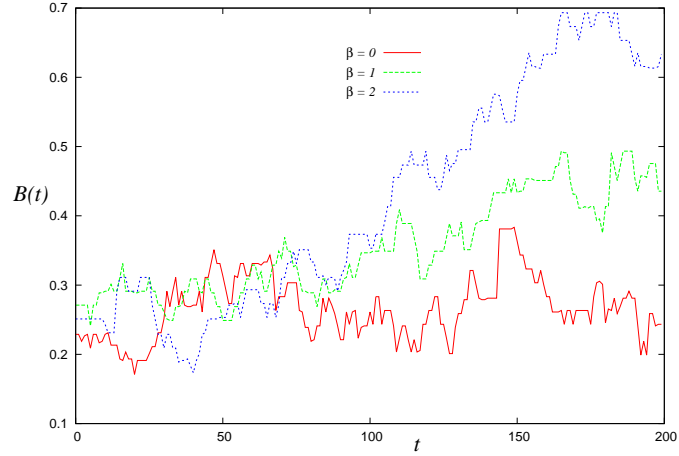}
\end{center}
\caption{\footnotesize 
The behavior of order parameter 
$B(t)$ which 
measures 
the degree of 
symmetry breaking 
in the $v_{k}$. 
The symmetry is apparently broken for 
$\beta=1,2$ 
around $t=100$ and 
the system changes from 
symmetric phase (small $B(t)$) to 
the symmetry breaking phase (large $B(t)$). 
}
\label{fig:fg65}
\end{figure}
We next evaluate the 
degree of the symmetry breaking by means of 
the following Shannon's entropy: 
\begin{equation}
H(t)=-\sum_{k=1}^{K}v_{k}(t) \log v_{k}(t)
\label{eq:entropy}
\end{equation}
where the above $H(t)$ takes the maximum 
for the symmetric solution $v_{k}(t)=1/K$ as 
\begin{equation}
H(t)=-K\times \frac{1}{K} \log (1/K) = \log K
\label{eq:entropy_max}
\end{equation}
whereas 
the minimum $H(t)=0$ is achieved for $v_k=1$ and 
$v_{k^{'} \neq k}=0$, which is apparently corresponding to 
the symmetry breaking phase. 
In Fig. \ref{fig:entropy}, 
we plot the $H(t)$ for several choices of $\beta$ as $\beta=0,1$ and $2$. 
From this figure, we find that 
for finite $\beta$, the system gradually moves from the symmetric 
phase to the symmetry breaking phase due to 
the endogenous information ({\it e.g.} word-of-mouth communication). 

\begin{figure}[ht]
\begin{center}
\includegraphics[width=10cm]{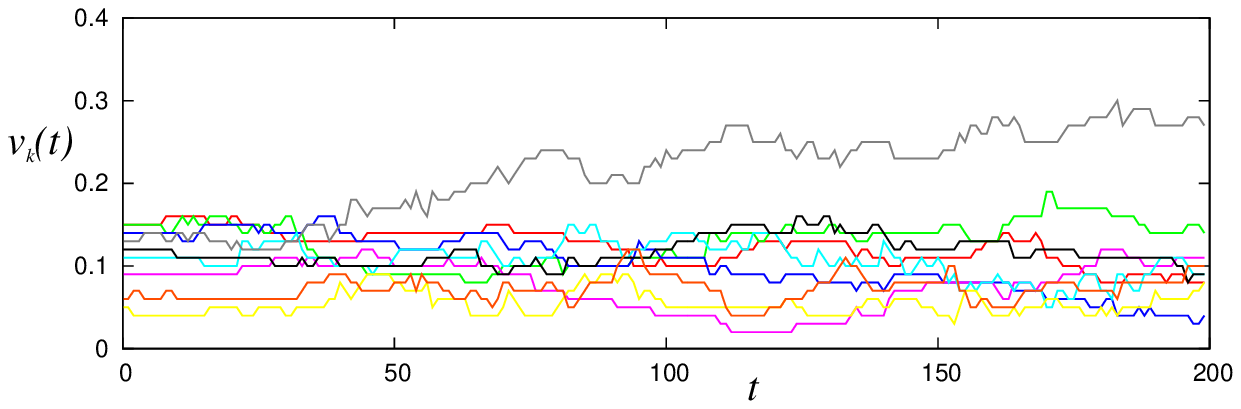} 
\includegraphics[width=10cm]{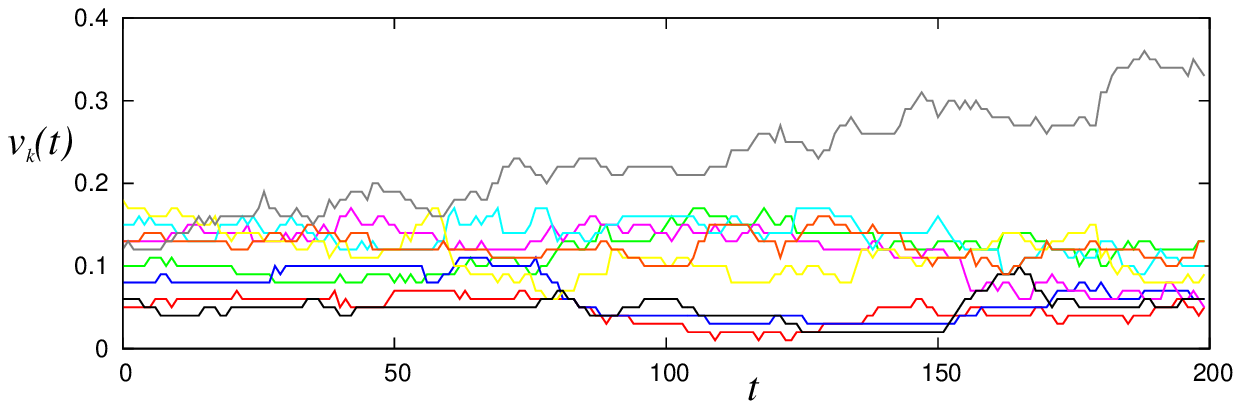}
\end{center}
\caption{\footnotesize 
Typical behavior of 
instant program rating point $v_{k}(t)$ for $k=1,\cdots,K$ 
for the choice of 
the time lag $r=10$ (the upper panel) and 
$r=40$ (the lower panel). 
Here we set $K=9, M=1, N=600$ and $T=200$. 
The parameters which specify the 
energy function of 
`lattice-type' controller are chosen as $(\gamma,\zeta,\beta)=(1,0,1.5)$ 
with history length $L=5$. 
The parameters appearing in the system are chosen as 
the same values as in Fig. \ref{fig:fg61}.}
\label{fig:fg65b}
\end{figure}

Finally, we consider the 
the time lag $r$-dependence (see equation (\ref{eq:lag})) of 
the resulting $v_{k}(t)$. 
The result is shown in Fig. \ref{fig:fg65b}. 
From this figure, we clearly find that 
the large time lag $r$ causes 
the large amount of symmetry breaking 
in the program rating point $v_{k}(t)$. 
\section{Adaptive location of commercials}
\label{sec:adapt}
In the subsection \ref{subsec:D}, 
we assumed that
each commercial advertisement is posted according to 
the Poisson process. 
However, it is rather artificial and 
we should consider the case in which each 
television station 
decides the location of 
the commercials using the adaptive manner. 
To treat such case mathematically, we 
simply set 
$K=2$ and $M=1$, 
namely, 
only two stations cast the same commercial 
of a single sponsor. 
Thus, we should notice that 
one can define 
$l_{k}(t)=0$ (on air) or 
$l_{k}(t)=1$ (CM) for $k=0,1$. 

Then, we assume that 
each television station decides 
the label $l_{k}(t)$ according to 
the following 
successive update rule of the CM location probability: 
\begin{eqnarray}
P(l_{k}(t)) & = &    
\frac{1}{4}
\left\{
1+ (2l_{k}(t)-1)
\tanh \Omega 
\left(
L_{c}-\sum_{\rho=t-1}^{t-\mathcal{L}}
l_{k}(\rho)
\right) 
\right\} \nonumber \\
\mbox{} \hspace{-0.5cm} & \times & 
\left\{
1+ (2l_{k}(t)-1)
\tanh 
\Omega 
\left(
\frac{v_{0}(t-\tau)-v_{0}(t - 1)}{\tau}
\right)
\right\} 
\end{eqnarray}
for $k=0,1$, 
which means that 
if the cumulative commercial time 
by the duration $\mathcal{L}$, that is, 
$\sum_{\rho=t-1}^{t-\mathcal{L}}
l_{k}(\rho)$ is lower than $L_{c}$, 
or if the slope of the program rating point $v_{k}$ 
during the interval $\tau$ is negative, 
the station $k$ is more likely to 
submit the commercial at time $t$. 
\begin{figure}[ht]
\begin{center}
\includegraphics[width=9cm]{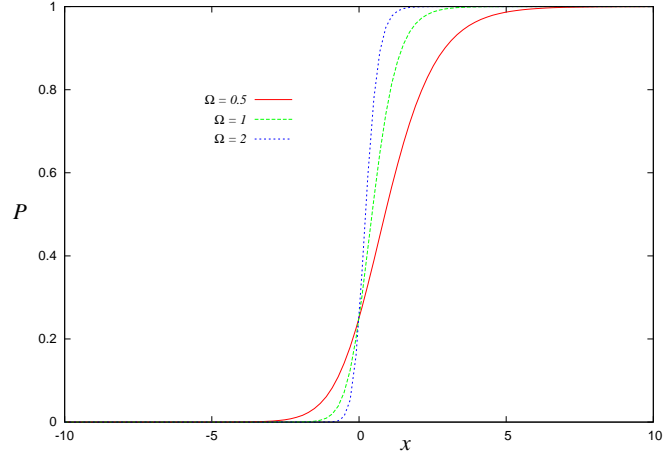}
\end{center}
\caption{\footnotesize 
The behavior of 
$P(x)=(1+\tanh(\Omega x))(1+\tanh(\Omega x))/4$. 
}
\label{fig:fg7}
\end{figure}
It should be noted that 
in the limit of $\Omega \to \infty$, 
the above probabilistic location 
becomes the following deterministic location model (see also Fig. \ref{fig:fg7}) 
\begin{equation}
l_{k}(t) = 
\Theta 
\left(
L_{c}-\sum_{\rho=t-1}^{t-\mathcal{L}}
l_{k}(\rho)
\right) 
\Theta 
\left(
\frac{v_{k}(t-\tau)-v_{k}(t - 1)}{\tau}
\right) 
\label{eq:lk}
\end{equation}
for $k=0,1$. 
From the nature of 
two television stations, 
$v_{0}(t-1)+v_{1}(t-1) = 1$ and 
$v_{0}(t-\tau)+v_{1}(t-\tau)  = 1$ 
should be satisfied. 
Thus, the possible combinations 
of $(l_{0}(t),l_{1}(t))$ are now restricted to 
$(l_{0}(t),l_{1}(t))=(0,0),(0,1),(1,0)$,  and 
$(l_{0}(t),l_{1}(t))=(1,1)$ is 
not allowed to be realized. 
In other words, for the deterministic 
location model described 
by (\ref{eq:lk}), 
there is no chance for the stations  
$k=0,1$ to cast the same CM advertisement at the same time. 

On the other hand, 
the viewer also might select the station according to 
the length of the commercial times in the past history. 
Taking into account the assumption, 
we define the station $\tilde{k}$ by 
\begin{equation}
\tilde{k} =  
\frac{1}{2}
\left\{
1 -{\rm sgn}(\langle l_{1}(t) \rangle - 
\langle l_{0}(t) \rangle)
\right\}
\end{equation}
with 
\begin{equation}
\langle l_{k}(t) \rangle  \equiv 
(1/\mathcal{L})
\sum_{\rho=t-1}^{t-\mathcal{L}}
l_{k}(t),\, \,\, k=0,1.
\end{equation} 
Then, 
the $\tilde{k}$ denotes 
the station which 
casted shorter commercial times 
during the past time steps $\mathcal{L}$ than the other. 
Hence, we rewrite 
the energy function 
for the two stations model 
in terms of the $\tilde{k}$ as follows. 
\begin{equation}
E_{k}(l) \equiv  - \zeta \prod_{\xi=s}^{e}\delta_{t_{\xi},t} \delta_{l, \overline{k}}- \beta \delta_{l,\hat{k}} 
-\xi \delta_{l,\tilde{k}}
\end{equation}
where we omitted the term 
$\gamma (\bm{Z}_{k}-\bm{Z}_{l})^{2}$ 
due to the symmetry $(\bm{Z}_{0}-\bm{Z}_{1})^{2}= 
(\bm{Z}_{1}-\bm{Z}_{0})^{2}$. 
As the result, the transition probability 
is rewritten as follows. 
\begin{equation}
P(l|k)  = 
\frac{{\exp}[
\zeta \prod_{\xi=s}^{e}\delta_{t_{\xi},t} \delta_{l, \overline{k}} + 
\beta \delta_{l,\hat{k}}
+\xi \delta_{l,\tilde{k}}
]}
{\sum_{l \neq k}
{\exp}[
\zeta \prod_{\xi=s}^{e}\delta_{t_{\xi},t} \delta_{l, \overline{k}} + 
\beta \delta_{l,\hat{k}} 
+\xi \delta_{l,\tilde{k}}
]} 
\end{equation}
The case without the exogenous information, 
that is, $\zeta=0$, 
we have the following simple transition probability for two stations. 
\begin{equation}
P(1|0) = 
\frac{{\exp}
(\beta \delta_{1,\hat{k}} + \xi \delta_{1,\tilde{k}})}
{{\exp}(\beta \delta_{0,\hat{k}} + \xi \delta_{0,\tilde{k}})
+
{\exp}
(\beta \delta_{1,\hat{k}} + \xi \delta_{1,\tilde{k}})} 
\label{eq:prob10}
\end{equation}
\begin{equation}
P(0|1)  = 
\frac{{\exp}
(\beta \delta_{0,\hat{k}} + \xi \delta_{0,\tilde{k}})}
{{\exp}(\beta \delta_{0,\hat{k}} + \xi \delta_{0,\tilde{k}})
+
{\exp}
(\beta \delta_{1,\hat{k}} + \xi \delta_{1,\tilde{k}})}
\label{eq:prob01}
\end{equation} 
with 
$P(0|0)=1-P(1|0)$ and 
$P(1|1)=1-P(0|1)$. 

\begin{figure}[ht]
\begin{center}
\includegraphics[width=10cm]{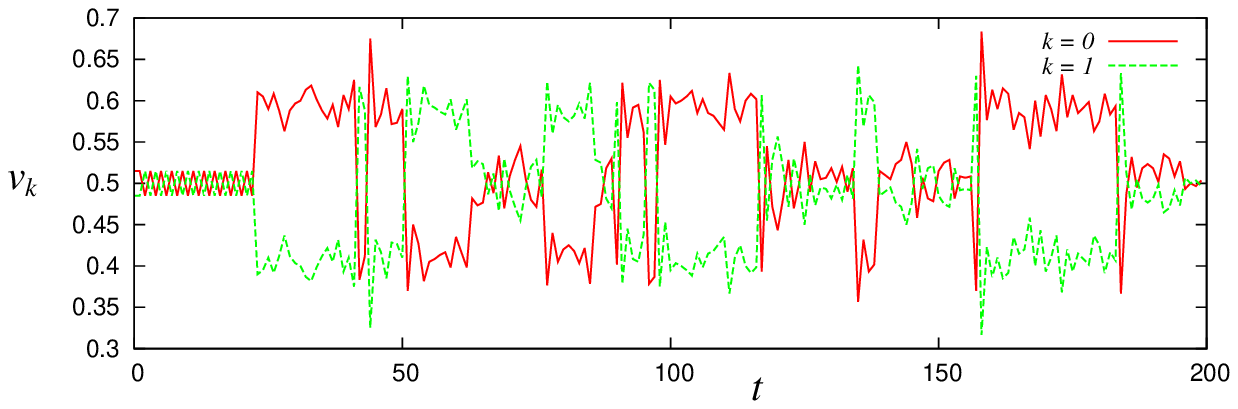} 
\includegraphics[width=10cm]{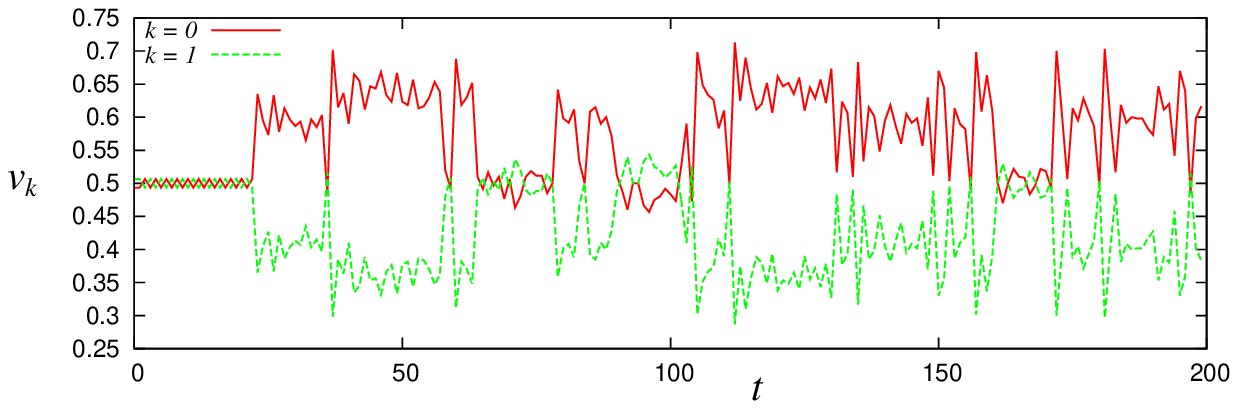}
\end{center}
\caption{\footnotesize 
Typical behavior of 
instant program rating points $v_{k}(t)$ for the two stations $k=0,1$ 
which are generated by the adaptive CM locations (\ref{eq:lk}) and 
zapping probabilities (\ref{eq:prob01})(\ref{eq:prob10}). 
We chose  $\beta=0$ (the upper panel) and 
$\beta=1$ (the lower panel). 
Here we set $M=1, N=600$ and $T=200$. 
The parameters which specify the 
energy function and CM locations are chosen as $\zeta=1, L_{c}=2, L=\mathcal{L}=\tau=20 ( =T/10)$. 
The other parameters appearing in the system are chosen as 
the same values as in Fig. \ref{fig:fg61}.}
\label{fig:fg8}
\end{figure}

We simulate the CM advertisement market described by 
(\ref{eq:prob10})(\ref{eq:prob01}) and (\ref{eq:lk}) 
and show the limited result in Fig. \ref{fig:fg8}. 
From this figure, we find 
that for the case of $\beta=0$, 
the superiority of two stations 
changes frequently, 
however, the superiority is almost `frozen', 
namely, the superiority does not change in time  
when we add the endogenous information 
to the system  
by setting $\beta=1$. 
For both cases ($\beta=0,1$), 
the behavior of the instant program rating as a `macroscopic quantity' seems to be 
`chaotic'. The detail analysis of this issue should be addressed as 
one of our future studies.  
\subsection{Frequent CM locations at the climax of program}
The results given in the previous sections 
partially have been reported by the present authors in the reference \cite{Kyan}. 
Here we consider a slightly different aspect of the television commercial markets. 

Recently in Japan, 
we sometimes have encountered 
the situation in which a television station 
broadcasts their CMs frequently at the climax of the program. 
Especially, in a quiz program, 
a question master speaks with an air of importance 
to open the answer and the successive CMs start before the answer comes out. 
Even after the program restarts,  
the master puts on airs and he never gives the answer and 
the program is again interrupted by the CMs. 
This kind of CMs is now refereed to as 
{\it Yamaba CM} (`Yamaba' has a meaning of `climax' in Japanese). 
To investigate the psychological effects on viewers' mind, 
Sakaki \cite{Sakaki} carried out 
a questionnaire survey and the result is given in Table \ref{tab:yamaCM}. 
\begin{table}[ht]
\begin{center}
\begin{tabular}{|l||c|c|c|}
\hline 
Question & Yes & I do not know & No \\
 \hline
 Is {\it Yamaba CM} unpleasant?  & 86 $\%$ & 7 $\%$ & 7 $\%$ \\
 \hline
Is {\it Yamaba CM} not favorable?  & 84 $\%$ & 14 $\%$ & 2 $\%$ \\
 \hline
 Do you purchase the product advertised by {\it Yamaba CM}? & 66 $\%$ & 37 $\%$ & 97 $\%$ \\
 \hline
 \end{tabular}
 \caption{\footnotesize 
 A questionnaire survey for viewers' impression on the so-called {\it Yamaba CM} \cite{Sakaki}.}
\label{tab:yamaCM}
 \end{center}
 \end{table}
From this table, we find that 
more than eighty percent of viewers might feel that 
the {\it Yamaba CM} is unpleasant and not favorable. 
With this empirical fact in mind, 
in following, we shall carry out computer simulations 
in which the CMs broadcasted by a specific television station 
are located intensively at the climax of the program. 
\subsubsection{Effects on the program rating points}
\begin{figure}[ht]
\begin{center}
\includegraphics[width=11cm]{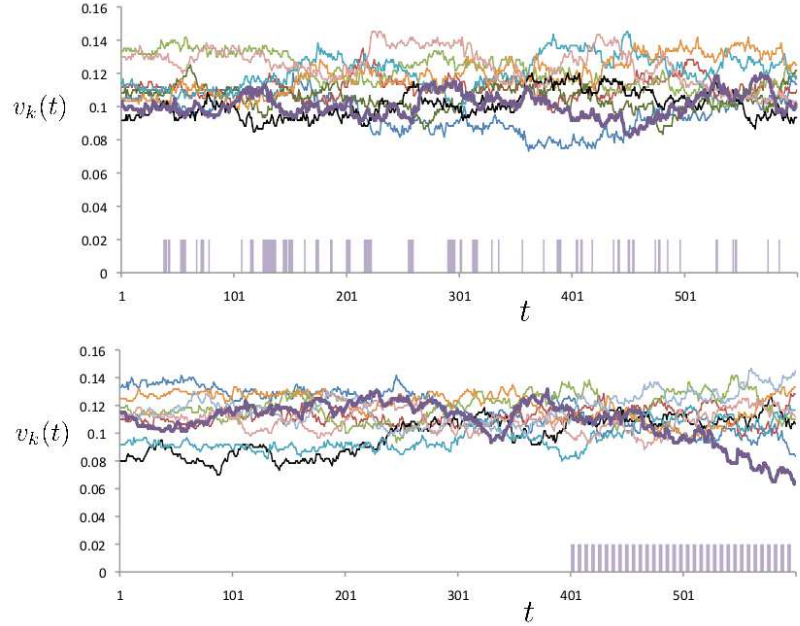}
\end{center}
\caption{\footnotesize 
The resulting program rating points for random locations of CMs (the upper panel) and 
for biased locations of CMs for a specific television station (say, $k=1$, the line in the panel is distinguished from the other eight stations by 
a purple thick line).  
The perpendicular purple lines stand for the period of CMs for the specific station $k=1$.}
\label{fig:yamaCM}
\end{figure}

We first consider the effects of the {\it Yamaba CMs} on the program rating points. 
The results are shown in Fig. \ref{fig:yamaCM} as 
a typical behavior of 
the program rating points $v_{k}(t),\,k=1,\cdots, 9$.
In this simulation, we fix the total length of 
CMs in a program so as to be less than eighteen percent of the total broadcasting time $T$ 
of the whole program including CMs. 
In the upper panel, 
we distribute the CMs of all television stations randomly, whereas in the lower panel, 
the CMs of a specific station (the line in the panel is distinguished from the other eight stations by 
a purple thick line)  
are located intensively at the climax (the end of the program) and 
for the other eight stations, the CMs are located randomly. 
The other conditions in the simulations are selected as the same as in Fig. \ref{fig:fg61}. 
From this figure, we find that 
the program rating point for the station 
$k=1$ which broadcasts {\it Yamaba CMs} intensively at the climax 
apparently decreases at the climax in comparison with the other stations. 

\begin{figure}[ht]
\begin{center}
\includegraphics[width=12cm]{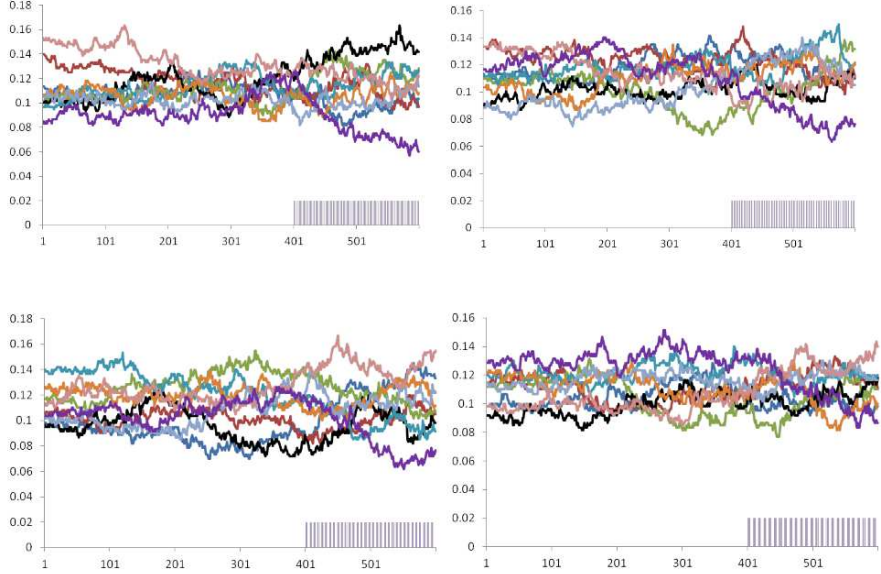}
\end{center}
\caption{\footnotesize
The $L_{\rm cm}$-dependence of the program rating points $v_{k}(t), \, k=1,\cdots,9$ 
shown in Fig. \ref{fig:yamaCM}. 
From the upper left to the lower right, 
$L_{\rm cm}=1, 2, 3$ and $L_{\rm cm}=5$. 
 }
\label{fig:yama_c_comp}
\end{figure}

To check the effect of the relaxation time 
$L_{\rm cm}$ on the results, 
we carry out the simulation by changing the 
value as $L_{\rm cm}=1,2,3$ and $5$.  
The results are shown in Fig. \ref{fig:yama_c_comp}. 
From this figure, we clearly find that 
the program rating point for the station 
$k=1$ which broadcasts {\it Yamaba CMs} intensively at the climax 
apparently decreases around $t=t_{\rm c} =400$ and 
the $t_{\rm c}$ is independent of 
the length of $L_{\rm cm}$. 
\subsubsection{Effects on the advertisement measurements}
\begin{figure}[ht]
\begin{center}
\includegraphics[width=12cm]{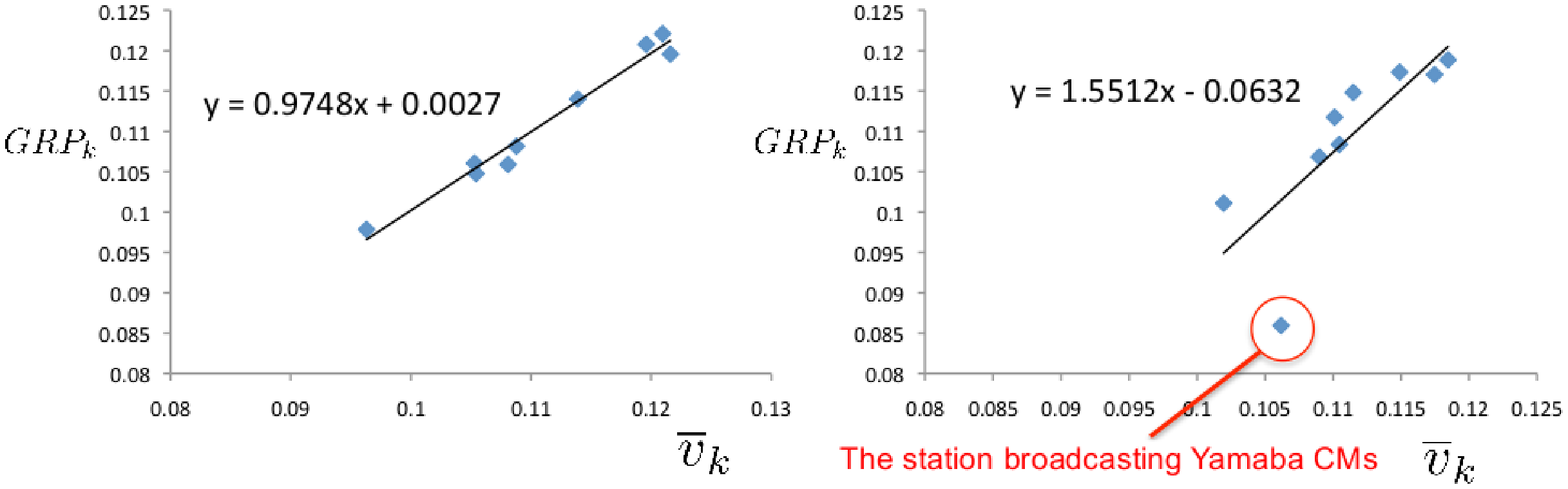}
\end{center}
\caption{\footnotesize 
The $GRP_{k}$-$\overline{v}_{k}$ 
diagram for $K=9$ stations. Each 
point corresponds to 
each station. 
In the left panel, there is no station broadcasting the so-called {\it Yamaba CMs}, whereas 
only a specific station, say $k=1$ is broadcasting the {\it Yamaba CMs} in 
the right panel. 
We set $M=1, N=T=600$. 
}
\label{fig:fg_ad1}
\end{figure}

We next evaluate of the effects of the so-called {\it Yamaba CMs} on 
the advertisement measurements such as the GRP or 
average contact time of the CMs by viewers. 
To quantify the effects, 
we consider the $GRP_{k}$-$\overline{v}_{k}$ diagram for $K=9$ stations, 
where $GRP_{k} \equiv GRP_{k}^{(1)}$ in the definition of 
(\ref{eq:rate3}) because now we consider the case of $M=1$ for simplicity. 
We plot the result in 
Fig. \ref{fig:fg_ad1} (left). 
In this panel, there is no station broadcasting the {\it Yamaba CMs}. 
When we define the advertisement efficiency $\eta$ for the sponsor 
by the slope of 
these points, the efficiency for this unbiased case is $\eta \simeq 0.9748$. 
We next consider the case 
in which a specific station, 
say, $k=1$ 
broadcasts the {\it Yamaba CMs}. 
The results are shown in the right panel of 
Fig. \ref{fig:fg_ad1}. 
From this panel, we are confirmed that 
the both $GRP_{k=1}$ and $\overline{v}_{k=1}$ apparently decrease 
in comparison with the other eight stations. 
Hence, the slope $\eta$ calculated by 
the eight stations (except for $k=1$) increases up to 
$\eta \simeq 1.5512$ because 
viewers who was watching the program of the station $k=1$ moved (changed the channel) 
to the other eight stations 
and it might increase the $GRP_{k \neq 1}$ and $\overline{v}_{k \neq 1}$ extensively. 

From the results given in this section, 
we might conclude that the {\it Yamaba CMs} (biased CM locations) are 
not effective from the view points of viewers, sponsors and 
television stations although our simulations were carried out for limited artificial situations. 
\section{Concluding remarks}
We proposed a `theoretical platform' 
to investigate the human collective behavior 
in the macroscopic scale through 
viewers' zapping actions at the microscopic level. 
We just showed  a very preliminary result without any comparison with 
empirical data. 
However, several issues, in particular, 
much more mathematically rigorous argument 
based on the queueing theory \cite{Inoue}, 
data visualization via the MDS \cite{Ibuki3}, 
portfolio optimization \cite{Livan}
and a mathematical relationship between 
our system and 
the so-called {\it regime-switching processes} \cite{Yin} 
should be addressed as our future studies. 
\subsection*{Acknowledgements}
This work was financially supported by 
Grant-in-Aid for Scientific Research (C) 
of Japan Society for 
the Promotion of Science, No. 22500195. 
One of the present authors (JI) thanks 
Nivedita Deo, Sanjay Jain, Sudhir Shah 
for fruitful discussion at the workshop:  {\it Exploring an Interface Between Economics $\&$ Physics} 
at University of Delhi. He also thanks George Yin at Department of Mathematics, Wayne State University, USA, 
for drawing our attention to the reference \cite{Yin}.  
We thank organizers of {\it Econophysics-Kolkata VII}, in particular, Frederic Abergel, Anirban Chakraborti, 
Asim K. Ghosh, Bikas K. Chakrabarti and Hideaki Aoyama. 


\begin{thebibliography}{99}

\bibitem{Reynolds}
C.W. Reynolds, 
{\it Flocks, Herds, and Schools: A Distributed Behavioral Model}, 
{\it Computer Graphics} {\bf 21}, 25 (1987). 


\bibitem{Makiguchi}
M. Makiguchi and J. Inoue, 
{\it Numerical Study on the Emergence of Anisotropy in Artificial Flocks: 
A BOIDS Modelling and Simulations of Empirical Findings},
{\it Proceedings of the Operational Research Society 
Simulation Workshop 2010 (SW10), CD-ROM}, pp. 96-102
 (the preprint version, arxiv:1004 3837) (2010). 
 See also 
 M. Makiguchi and J. Inoue, 
 {\it Emergence of Anisotropy in Flock Simulations and Its 
Computational Analysis}, 
 {\it Transactions of the Society of Instrument and Control Engineers} {\bf 46}, No. 11, 
 pp. 666-675 (2010) (in Japanese). 
 
 \bibitem{Ibuki}
T. Ibuki and J. Inoue, 
{\it Response of double-auction markets to instantaneous Selling-Buying signals with stochastic Bid-Ask spread}, 
{\it Journal of Economic Interaction and Coordination} {\bf 6}, No.2, pp. 93-120 (2011). 

 
 \bibitem{Siddarth}
 S. Siddarth and A. Chattopadhyay, 
 {\it To Zap or Not to Zap: 
 A Mixture Model of Channel Switching 
 During Commercials}, 
 Working Paper, Series No. MKTG97.091 (1997). 
 
 \bibitem{Ikai}
 H. Ohnishi, K. Ishida, H. Aoyama, 
 Y. Saruwatari and M. Ikai, 
 {\it The Operations Research Society of Japan} 
 {\bf 50}, No. 3, pp. 151-158 (2005) (in Japanese). 
 


\bibitem{Chen}
H. Chen and J. Inoue, 
{\it Dynamics of probabilistic labor markets: statistical physics perspective}, 
{\it  Lecture Notes in Economics and Mathematical Systems} {\bf 662}, pp. 53-64, ``Managing Market Complexity", 
Springer (2012), 
H. Chen and J. Inoue, 
{\it Statistical Mechanics of Labor Markets}, 
{\it Econophysics of systemic risk and network dynamics, New Economic Windows}, Springer-Verlag (Italy-Milan), 
pp. 157-171 (2013).

\bibitem{Ibuki2}
T. Ibuki, S. Higano, S. Suzuki and J. Inoue, 
{\it Hierarchical information cascade: visualization and prediction of human collective behaviour at financial crisis by using stock-correlation}, 
{\it ASE Human Journal} {\bf 1}, Issue 2, pp. 74-87 (2012).


\bibitem{Kyan}
H. Kyan and J. Inoue, 
{\it Modeling television commercial advertisement markets: To zap or not
to zap, that is the question for sponsors of TV programs}, 
{\it Proceedings of IEEE/SICE Symposium SII2012}, CD-ROM, pp. 816-823 (2012). 


\bibitem{Sakaki}
H. Sakaki, 
{\it Annual Report of 
Nikkei Advertising Research Institute} {\bf 255}, pp. 19-26 (2011). 



\bibitem{Inoue}
N. Sazuka, J. Inoue and E. Scalas, 
{\it The distribution of first-passage times and durations in FOREX and future markets}, 
{\it Physica A} {\bf 388}, No. 14, pp. 2839-2853 (2009), 
J. Inoue and N. Sazuka, 
{\it Queueing theoretical analysis of foreign currency exchange rates}, {\it Quantitative Finance}  {\bf 10}, No. 10, pp. 121-130 (2010). 

\bibitem{Ibuki3}
T. Ibuki, S. Suzuki and J. Inoue, 
{\it  Cluster Analysis and Gaussian Mixture Estimation of 
Correlated Time-Series by Means of Multi-dimensional Scaling}, 
{\it Econophysics of systemic risk and network dynamics, New Economic Windows}, Springer-Verlag (Italy-Milan), pp. 239-259 (2013).


\bibitem{Livan}
G. Livan, J. Inoue and E. Scalas, 
{\it On the non-stationarity of financial time series: impact on optimal portfolio selection}, 
 {\it Journal of Statistical Mechanics}, P07025 (2012).

\bibitem{Yin}
S.L. Nguyen and G. Yin, 
{\it Weak convergence of Markov-modulated random sequences}, 
{\it Stochastics: An International Journal of Probability and Stochastic
Processes} {\bf 82}, No. 6, pp. 521-552 (2010). 
\end{thebibliography}
\end{document}